# Pre-treatment of outliers and anomalies in plant data: Methodology and case study of a Vacuum Distillation Unit

Kamil Oster[a, b*], Stefan Güttel[a*], Jonathan L. Shapiro[c], Lu Chen[b*], Megan Jobson[d]

[a] Department of Mathematics, The University of Manchester, Alan Turing Building, Oxford Road, Manchester, M13 9PL, UK

[b] Process Integration Limited, Station House, Stamford New Road, Altrincham, WA14 1EP, UK

[c] Department of Computer Science, The University of Manchester, Kilburn Building, Oxford Road, Manchester, M13 9PL, UK

[d] Department of Chemical Engineering and Analytical Science, The University of Manchester, The Mill, Sackville Street, Manchester, M1 3AL, UK

* kamil.oster@manchester.ac.uk, kamil.oster@processint.com, stefan.guettel@manchester.ac.uk, lu.chen@processint.com

**Abstract**

Data pre-treatment plays a significant role in improving data quality, thus allowing extraction of accurate information from raw data. One of the data pre-treatment techniques commonly used is outliers detection. The so-called $3\sigma$ method is a common practice to identify the outliers (using triple standard deviation as upper and lower limits). However, as shown in the manuscript, it does not identify all outliers, resulting in possible distortion of the overall statistics of the data. This problem can have a significant impact on further data analysis and can lead to reduction in the accuracy of predictive models. There is a plethora of various techniques for outliers detection, however, aside from theoretical work, they all require case study work.

In this work, two types of outliers were considered: short-term (erroneous data, noise) and long-term outliers (mainly malfunctioning for longer periods of time). The data used were taken from the vacuum distillation unit (VDU) of an Asian refinery and included data points from 40 physical sensors (temperature, pressure and flow rate). We used a modified method for $3\sigma$ thresholds to identify the short-term outliers. More specifically, sensors data are divided into chunks determined by change points and $3\sigma$ thresholds are calculated within each chunk representing near-normal distribution. We have shown that piece-wise $3\sigma$ method offers a better approach to short-term outliers detection than $3\sigma$ method applied to the entire time series. Nevertheless, this does not perform well for long-term outliers (which can represent another state in the data). In this case, we used principal component analysis (PCA) with Hotelling's $T^2$ statistics to identify the long-term outliers. The results obtained with PCA were subject to DBSCAN clustering method. The outliers (which were visually obvious and correctly detected by the PCA method) were also correctly identified by DBSCAN which supported the consistency and accuracy of the PCA method.

## 1. Introduction

In most applications making use of historic data, data pre-treatment is needed to ensure quality. One of the steps in data pre-treatment is identification and removal or correction of outliers in the data. Hawkins (1980) defines an outlier as '*an observation which deviates so much from other observations as to arouse suspicions that it was generated by a different mechanism*'.[1] Outlier correction is important because outliers may alter the distribution of data points in datasets, potentially distorting findings from data analysis and reducing the accuracy of predictive models. Blázquez-García *et al.* (**2021**) performed a state-of-the-art review of outlier/anomaly detection in time series, particularly a taxonomy of outliers.[2] Two classes of outliers of particular interest in this work are point and subsequence outliers. Point outliers are usually noise or erroneous data which behave unusually in a specific time instant, particularly when compared to neighbouring records.[3] Subsequence outliers refer to consecutive points in time whose joint behaviour is unusual.

Methods for identifying outliers must take into account the type and properties of the data at hand. Generally, the available methods can be classified into two categories: univariate and multivariate outliers detection. Univariate detection methods consider a single, time-dependent variable, while multivariate detection methods work with more than one time-dependent variable simultaneously. Some of the techniques for univariate outlier detection are based on constant or piece-wise models (applying basic statistical measures such as standard deviation, median or median absolute deviation),[4,5] B-splines,[6] exponentially weighted moving average,[7] Gaussian mixture models,[8] or autoregressive integrated moving average.[9] Some of the multivariate techniques make use of long short-term memory neural networks (LSTMs),[10] or principal component analysis (PCA).[11] Data can have varied sources which makes it challenging to create a universal method to determine outliers, thus, case studies provide information on suitable methods for certain types of data.

The data used in this work are taken from the vacuum distillation unit (VDU) of a petroleum refinery operating in China. This example consists of 40 physical sensors of 3 types: temperature (23 sensors), pressure (2 sensors) and flow rate (15 sensors), as shown in **Figure 1**. Two types of outliers are typical in data of this sort. Firstly, there might be outliers related to noise or erroneous records which include point and obvious outliers, so-called short-term outliers. Secondly, outliers might arise from the malfunctioning sensors that deviate from the overall trend, so-called long-term outliers. A literature review of data with similar origins showed that many works either use the $3\sigma$ method (described in the *2. Methods* section) for outliers detection,[12–14] or do not pre-treat their data at all.[15–17] For example, Rogina *et al.* (2011) performed statistical analysis, including upper and lower quartile of the data sets,[18] and showed that there were records below/above the lower/upper quartile, respectively. However, the authors decided not to treat the outliers (for example imputation or correction).

This work focuses on the VDU case study, in particular identification of short- and long-term outliers. The *2. Methods* section provides a justification for the studies carried out in this work and theoretical background for techniques used. *3. Results and Discussion: Case Study* section is divided into two parts: short-term outliers detection and long-term outliers detection. For short-term outliers, we applied the $3\sigma$ method to entire time series and a piece-wise approximated $3\sigma$ method, coupled with change point detection. For long-term outliers, we applied PCA to identify the long-term outliers coupled with DBSCAN clustering.

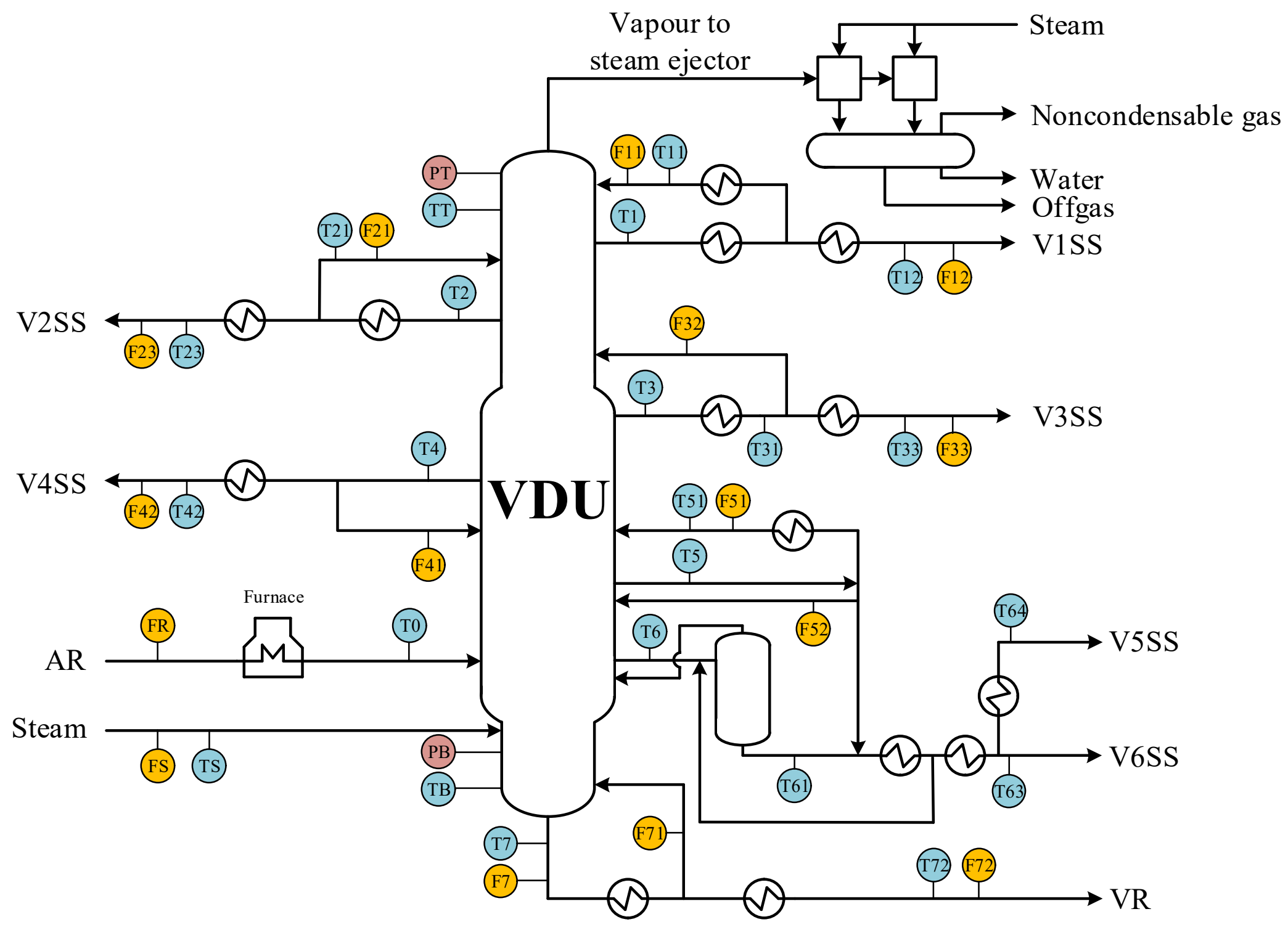

**Figure 1.** Flow Diagram of Vacuum Distillation Unit (VDU) showing sensors.

## 2. Methods

### 2.1. Statement of the problem

Data pre-treatment is an important step towards extracting useful information from data, particularly those collected from chemical processing plants. For example, statistical inference based on mean and standard deviation is a common approach to measure and control the fluctuation of an industrial process in this context, but the presence of outliers can significantly distort the mean response.[19] This distortion can have negative consequences for process control, particularly when the number of data samples is small and the mean is sensitive to outliers. A population central tendency can be estimated well through the sample mean at which one outlier is negligible compared to a substantial sample population.[20] The median is a more reliable descriptor of population central tendency.

There are different potential sources of outliers in chemical plant data. In this work, two types of outliers, presented in **Figure 2**, are of relevance. Short-term outliers are sporadic outliers that are caused by measurement errors, such as random fluctuations of the signal or sudden events in the process detected by the sensor (**Figure 2a**). These usually manifest as single, isolated, data points deviating from the overall statistics. Long-term outliers can originate from faulty or malfunctioning sensors. Unlike short-term outliers, long-term outliers span over longer periods such as hours or days. An example of long-term outlier was shown in **Figure 2b**. Additionally, a short-term outlier is present within the long-term outlier shown.

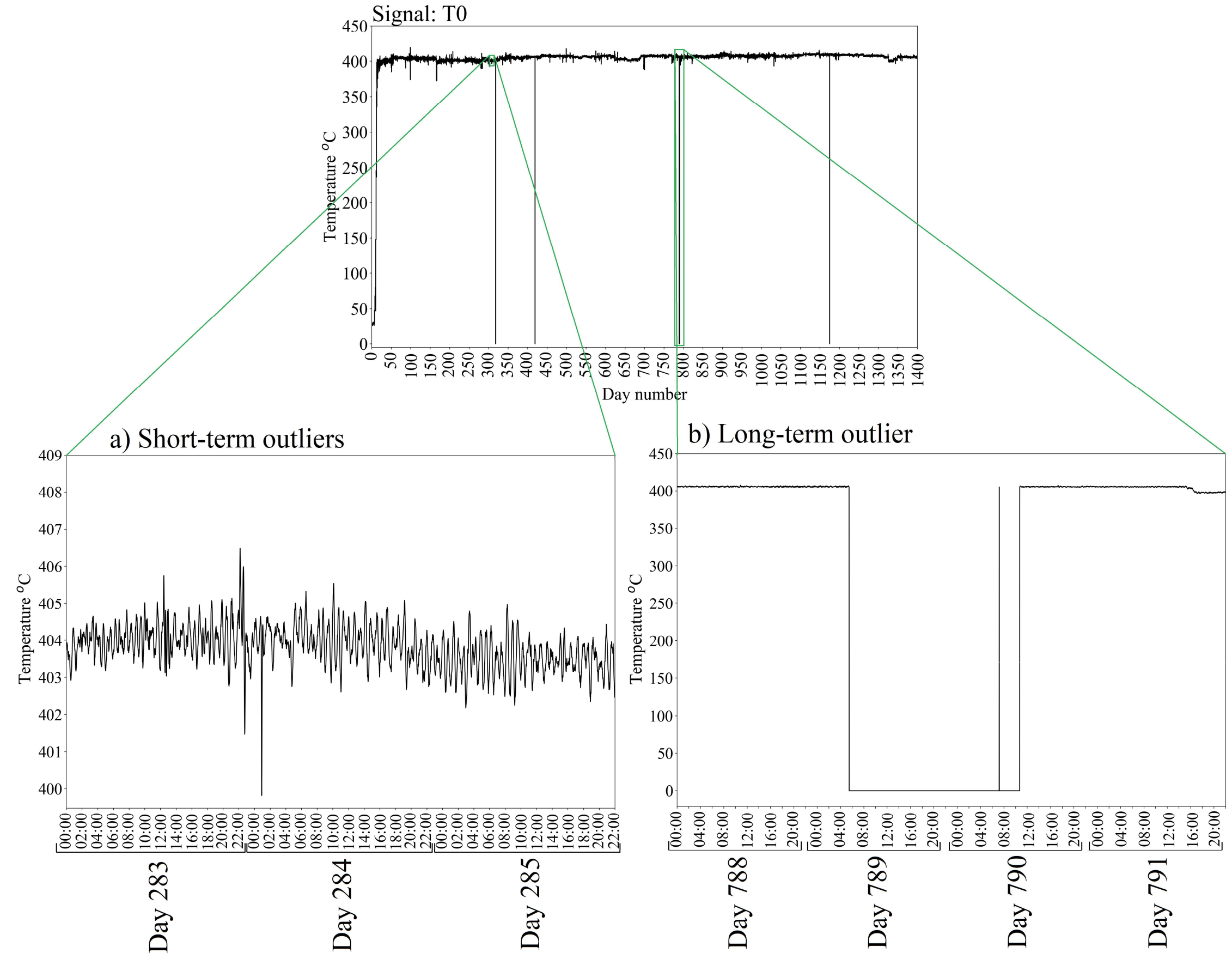

**Figure 2.** An example of (**a**) short-term and (**b**) long-term outliers (short-term outlier present within the long-term outlier).

The most common method to determine outliers in data is the so-called $3\sigma$ method.[21] This method uses the three-fold standard deviation or median absolute deviation as upper and lower limits for the data, *i.e.* any data above or below these limits, respectively, are classed as outliers. The major drawback of this method is that it relies on standard deviation of the entire time series which can be heavily influenced by the outliers themselves (especially long-term outliers).[22] We discuss this issue in more detail in the *3. Results and Discussion: Case Study* section below. Other significant effects are natural fluctuations of the signals as an effect of changes in the chemical process, or changes in the level of noise. The mean or median used as the baseline in the $3\sigma$ method can be heavily skewed by these factors, producing over- or under-estimated upper and lower limits. While it may be possible to identify the long-term outliers (such as in **Figure 2b**) with the $3\sigma$ method, other less obvious outliers (such as in **Figure 2a** or short-term outlier present within the long-term outlier in **Figure 2b**) are undetectable.

### 2.2.Short-term outliers

Our analysis initially focuses on short-term outliers. The $3\sigma$ method is modified so that instead of applying it on the entire time series, the data is segmented into smaller pieces with near-normal distribution of data on each of which the $3\sigma$ method is applied separately. The start and end points for the pieces are considered as change points and, hence, our first concern is how to determine these change points.

Several change point detection methods exist.[23–26] They are classified into supervised methods (*e.g.* multi-class classifiers, binary class classifiers and virtual classifiers, with techniques such as decision tree,[27] nearest neighbour,[28] support vector machine,[27] Naïve Bayes,[29] and hidden Markov model[29]), and unsupervised methods (*e.g.* likelihood ratio,[30,31] subspace model,[32] probabilistic methods,[33] kernel based methods,[34] graph based methods,[35] and clustering[36]). Methods based on the likelihood ratio were shown to be a good approach for detecting both single and multiple change points in the entire time series, with reasonable performance *vs* computational cost trade-off.[37–39] Off-line change point detection methods assume that the entire data sequence is available for the analysis. On the contrary, on-line change point detection analyses an incoming data stream in real-time to identify the change points.

A ready-to-use algorithm incorporating likelihood ratio methods for change point analysis is available in the R programming language in the '*changepoint*' package.[40] It offers a lot of versatility for the user, *e.g.* regarding the choice of cost and penalty functions, distribution statistics, as well as number of change points to be determined. For this work, we have adapted the '*cpt.mean*' module of the '*changepoint*' package in Python. Its principles are described herein.

We consider a one-dimensional time series, $t_{1:N} = (t_1, \dots, t_N)$ with a change point occurring at a time index $\tau \in \{1, \dots, N-1\}$, resulting in two subsets of $\{t_1, \dots t_\tau\}$ *and* $\{t_{\tau+1}, \dots, t_N\}$ with statistical properties $\theta_{0:\tau}(\mu_{1:\tau}, \sigma^2_{1:\tau})$ and $\theta_{\tau:N}(\mu_{\tau:N}, \sigma^2_{\tau:N})$, respectively, where $\mu$ is the mean and $\sigma$ is the standard deviation of the sample. Two scenarios are considered:

- Null hypothesis (with no change point occurring) $H_0: \theta_0 = \theta_1 = \cdots = \theta_{N-1} = \theta_N$,

and,

- Alternative hypothesis (with change point occurring) $H_1: \theta_0 = \theta_1 = \cdots = \theta_{\tau-1} = \theta_\tau \neq \theta_{\tau+1} = \theta_{\tau+2} = \cdots \theta_{N-1} = \theta_N$.

Under the assumption that the data is sampled from a known distribution, a normal distribution in our case, the likelihood of observing the data under the hypotheses $H_0$ and $H_1$ can be calculated as follows:

$$H_0: \mathcal{L}(H_0) = p(x|H_0) = \prod_{i=t_0}^{N} p(x_i|\theta_0) \qquad (1)$$

$$H_1: \mathcal{L}(H_1) = p(x|H_1) = \prod_{i=t_0}^{\tau} p(x_i|\theta_1) \prod_{j=\tau+1}^{N} p(x_j|\theta_2) \quad (2)$$

where $\mathcal{L}$ is likelihood. A higher likelihood $\mathcal{L}(H_0)$ indicates a stronger $H_0$ hypothesis and a less probable change point: To compare the two likelihoods, it is convenient to work with the log-likelihood ratio:

$$\mathcal{R}_\tau = \log\left(\frac{\mathcal{L}_{H_1}}{\mathcal{L}_{H_0}}\right) = \sum_{i=1}^{\tau} \log p(x_i|\theta_1) + \sum_{j=\tau+1}^{N} \log p(x_j|\theta_2) - \sum_{k=1}^{N} \log p(x_k|\theta_0) \quad (3)$$

The log-likelihood ratio can be employed to search for time instances where change points occur. A generalised log-likelihood ratio $G$ can be defined as the maximum of $\mathcal{R}_\tau$ for all possible $\tau$:

$$G = \max_{1 \leq \tau \leq N} \mathcal{R}_\tau \quad (4)$$

In what follows, we assume that the time series data within each of the pieces are normally distributed:

$$f(x|\mu, \sigma) = \frac{1}{\sqrt{2\pi\sigma^2}} e^{-(x-\mu)^2/2\sigma^2} \quad (5)$$

In that instance, a change point can correspond to a change in mean, $\mu$, or change in variance, $\sigma^2$. In this manuscript, we only consider changes in mean as these are the most natural consequences of changes in a chemical process. Thus,

- Null hypothesis (no change point): $\mathcal{L}(H_0) = \left(\frac{1}{\sqrt{2\pi\sigma^2}}\right)^N \prod_{i=1}^{N} \exp\left[-\frac{(x_i-\mu_0)^2}{2\sigma^2}\right]$

and,

- Alternative hypothesis with change point at $\tau$:

$$\mathcal{L}(H_1) = \left(\frac{1}{\sqrt{2\pi\sigma^2}}\right)^N \prod_{i=1}^{\tau} \exp\left[-\frac{(x_i-\mu_1)^2}{2\sigma^2}\right] \prod_{j=\tau+1}^{N} \exp\left[-\frac{\left(x_j-\mu_2\right)^2}{2\sigma^2}\right]$$

Hence, $\mathcal{R}_\tau$ is reduced to the following three-term equation:

$$\mathcal{R}_\tau = \log\left(\frac{\mathcal{L}_{H_1}}{\mathcal{L}_{H_0}}\right) \propto -\frac{1}{2\sigma^2}\left[\sum_{i=1}^{\tau}(x_i-\mu_1)^2 - \sum_{j=\tau+1}^{N}\left(x_j-\mu_2\right)^2 + \sum_{k=1}^{N}(x_k-\mu_0)^2\right] \quad (6)$$

An example of this approach applied to a selected signal from the VDU is presented in **Figure 3**. **Equation 6** is applied to the entire time series with varying $\tau \in \{1, \ldots, N-1\}$, producing a sequence of $\mathcal{R}_\tau$ values (green line in **Figure 3**). The global maximum (associated with **Equation 4**) determines the position of a single change point at $t = 3545$.

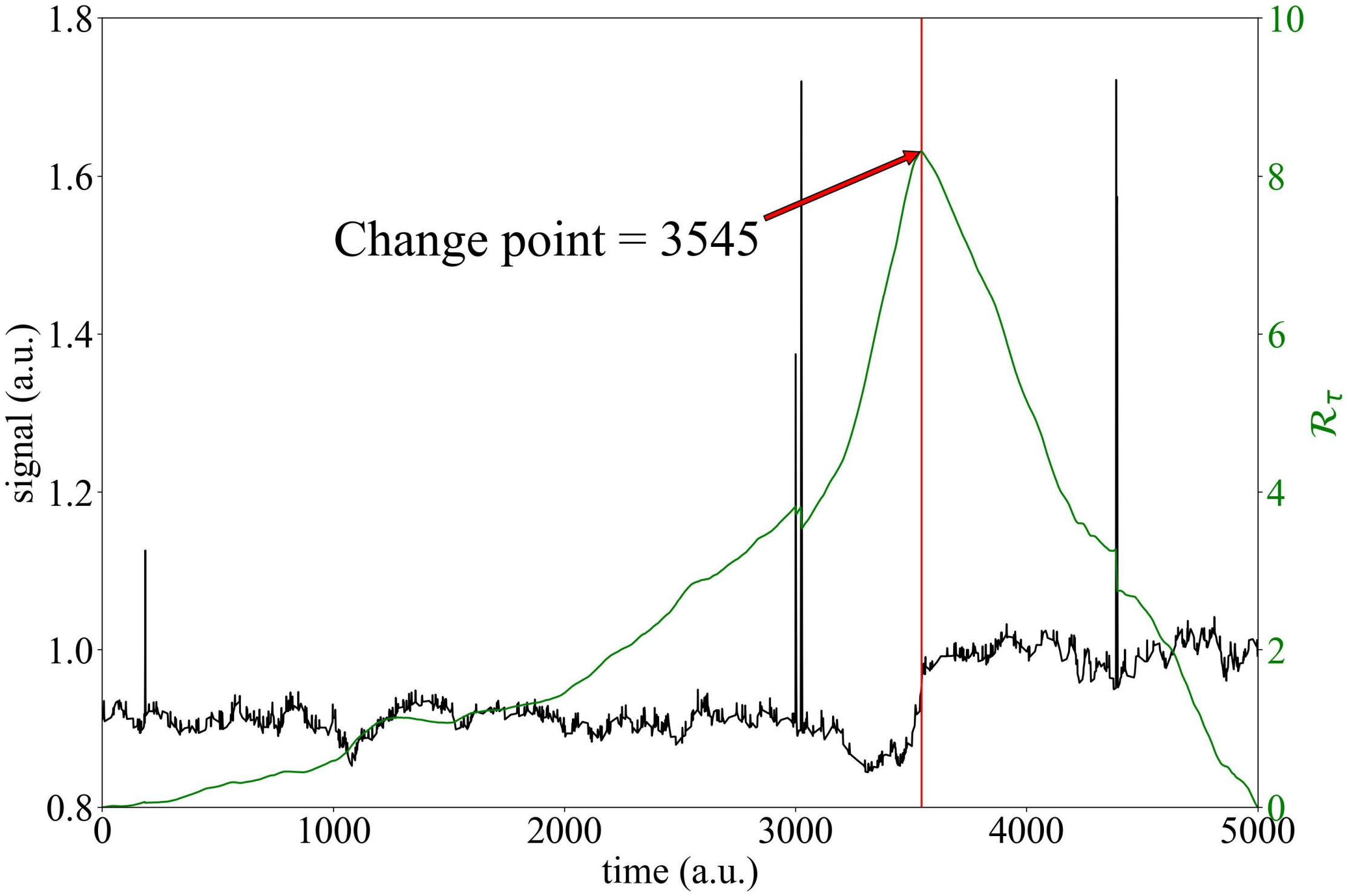

**Figure 3.** An example of detecting single change point; black line – original time series, green line – calculated $\mathcal{R}_\tau$, red vertical line – detected change point.

The approach described above can also be applied to detect multiple change points online where $\mathcal{R}_\tau$ is calculated with more data added with a sliding window. The principles are similar to those of Adaptive sliding WINdow (ADWIN).[41] Here we summarise the key concepts, with **Figure 4** showing a pseudocode of the algorithm. ADWIN uses a variable length sliding window of recent observations and discard older records in the window upon the evidence that these older observations vary significantly from the rest of the window. Herein, we use a threshold at which the maximal value *G* indicates a change point. A number of various information criteria were studied in the literature, for example *AIC*, *MAICE*, *SIC*, *BIC*.[26] However, these have to be optimised against the properties of the time series under consideration, which is a drawback. The data studied in this manuscript cover a range of physical quantities, including flow rate, temperature and pressure. As discussed in later parts of this manuscript, they vary in their statistics, and this makes it impractical to use a single information criterion. Hence, a less constraining parameter is introduced, namely the minimum length of a piece $L_{min}$. A candidate for a change point is again identified *via* an extremum (as *G* from **Equation 4**). Starting from the position of a potential change point (change in the sign of d*R*/d*t*), it is assumed that for the length of $L_{min}$ (*e.g.* in case of data with frequency of 1 minute, 1 hour window is $L_{min}$=60), no other maxima are present. Secondly, we define the number of data points added incrementally in each iteration (also related to the time step and sampling resolution) is denoted as *n*.

The search for change points starts with taking the first *n* data points (step 1 in **Figure 4**) to calculate the $\theta_0$ statistics (*i.e.* $\mu_0$). After that, the next *n* data points are taken (step 2 in **Figure 4**). $\theta_1$ statistics (*i.e.* $\mu_1$) are calculated for the added data, and $\theta_2$ statistics (*i.e.* $\mu_2$) are calculated for all data (either from starting point or the last identified change point). **Equation 6** is used to evaluate $\mathcal{R}_\tau$ (step 3 in **Figure 4**). This procedure is iterated until a function of $\mathcal{R}_\tau$ reaches G (a candidate for a change point, step 4 in **Figure 4**). From this point, the iterations of the algorithm (steps 2-4) check whether the condition of $L_{min}$ is met: **a)** if no other *G* is detected within $L_{min}$, the potential change point is recorded and the algorithm starts over from step 1, where $\mu_0$ begins at that change point; **b)** if a new maximal value *G* was detected, the latest *G* becomes a new candidate for a change point, and iterations continue.

| |
|---|
| **1. Initialise**<br>Take the first $n$ observations for $\mu_0$<br><br>**2. New datum**<br>Take the next $n$ observations to calculate $\mu_1$ and $\mu_2$<br><br>**3. Evaluate $R_t$**<br>Use equation 6 using the above $\mu_0$ , $\mu_1$ and $\mu_2$<br><br>**4. Repeat step 2 until extremum observed and check for $L_{min}$**<br>Calculate new $R_t$ for the condition in equation 4, and then continue for the piece equal to $L_{min}$<br><br>**If:** No new $G$ from equation 4 was observed within $L_{min}$ – time at step 4 is a changepoint + replace $\mu_0$ with a value at new changepoint $\tau = t_n$, so $\mu_0 = \mu_n$<br><br>**If:** New $G$ from equation 4 was observed within $L_{min}$ – continue scanning until a new condition in step 4 |

**Arguments:**

| | |
|---|---|
| ***n*:** | (int) Time step for scanning (or resolution), *e.g.* n = 10 means taking datum every 10th point into the distribution (resolution) |
| **$L_{min}$:** | (int) Minimum length of the piece between two changepoints |

**Figure 4.** Algorithm for on-line multiple change point search.

The mean-based change point detection described above was applied to the same signal as discussed above, and the results are shown in **Figure 5** ($n$ = 10, $L_{min}$ = 60). Unfortunately, the ground truth of change points is not available in our application, *i.e.* it is not possible to assess whether the change points detected are associated with actual changes in the chemical process that may result in changes in the signal. Nevertheless, the detected change points can serve as an indicator of significant variations in the data.

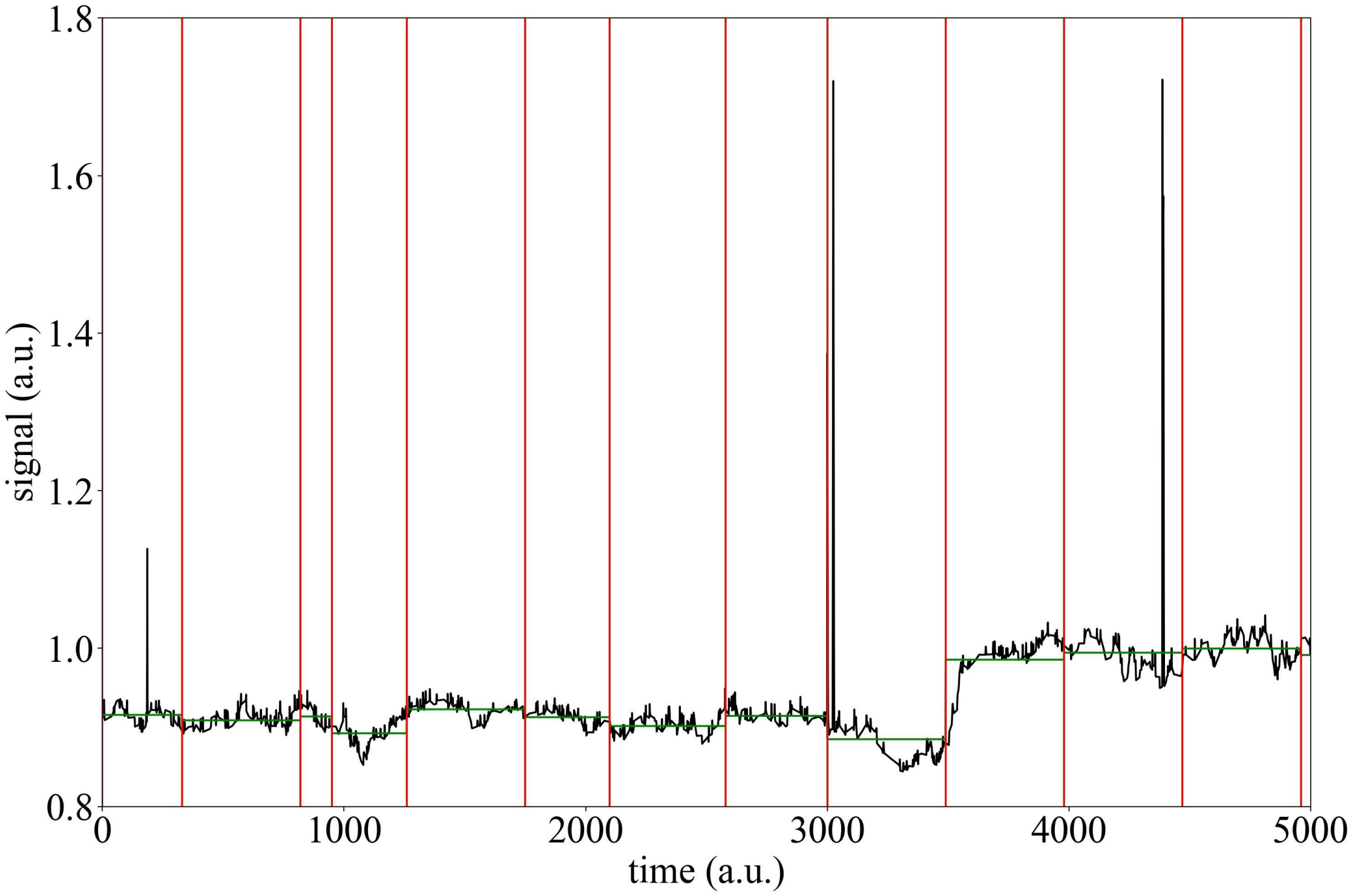

**Figure 5.** Detected change points and estimated pieces through on-line change points search approach (as depicted in **Figure 4**); black line – original time series, green line – estimated runs, red vertical lines – detected change points.

### 2.3. Long-term outliers

Principal component analysis (PCA) was used to identify long-term outliers. All sensors' data are scaled with zero-mean and unit-variance. Principal component analysis (PCA) decomposes the historical data matrix $X \in \mathbb{R}^{N \times M}$ with $N$ as length of the dataset and $M$ as number of sensors, the columns of which we assume to have zero mean and unit norm:

$$X^T X P_i = \lambda_i P_i \quad (8)$$

so that:[20,42]

$$X = T_1 P_1^T + T_2 P_2^T + \cdots + T_A P_A^T + E \quad (7)$$

Here, each $T_i \in \mathbb{R}^N$ is a score vector, $E \in \mathbb{R}^{N \times M}$ is a residual matrix of minimal norm, and $P_i \in \mathbb{R}^M$ are orthonormal vectors obtained as the eigenvectors of the covariance of matrix $X$ with associated eigenvalues $\lambda_1 \geq \lambda_2 \geq \cdots \geq \lambda_A$. The score vectors are obtained as

$$T_i = XP_i \quad (9)$$

Typically, $A \leq M \leq N$ and often just a small number of principal components $A$ is retained (such as $A = 2$). An approximation to $X$ using PCA is obtained by simply dropping the residual matrix:

$$\hat{X} = T_1 P_1^T + T_2 P_2^T + \cdots + T_A P_A^T \quad (10)$$

PCA is a popular tool to visualise high-dimensional data sets, using the $i$-th entries of the score vectors as coordinates to represent the $i$-th row of the data matrix. This projection approach can highlight outliers; hence, it is referred to as an outlier map. Practically, the first two principal components are used to create such visualisation.[43,44] In this manner, an outlier is defined as a data point far from the data in subspace spanned by the two eigenvectors.

A measure to quantify how much a point deviates from the overall data set, Hotelling's $T^2$ statistics can be used.[45] $T^2$ statistics can be used to detect periods during which sensor faults occurred. Generally, Hotelling's $T^2$ measures the variation of each sample in the PCA model:

$$T^2(i) = t_i \Lambda^{-1} t_i^T \quad (11)$$

where $\Lambda = \mathrm{diag}[\lambda_1, \lambda_2, \ldots, \lambda_A]$ and $t_i$ is $i$-th observation of $[T_1, T_2, \ldots, T_A]$. A threshold for $T^2$ can be calculated by using $F$ distribution at $(1 - \alpha)$ confidence level:

$$T_\alpha = \frac{(M^2-1)A}{M(M-A)} F_\alpha(A, M-A) \quad (12)$$

where $F_\alpha(A, M - A)$ is a critical value with $A$ and $M - A$ degrees of freedom. Time points at which $T^2(m) > T_\alpha$ are classed as outliers.

The outlier map constructed based on the first two principal components should produce at least two clusters: 1) the main cluster which contains points within the confidence threshold, 2) other clusters which are related to outliers. The main cluster should have a centroid at (0, 0), assuming the sensors data were zero-mean and unit-variance centred before using PCA.

In order to identify clusters with little to no supervision, we use the DBSCAN algorithm.[46] As will be seen in the *3. Results and Discussion: Case Study* section below, this density-based algorithm is well suited for this task as the points corresponding to the first two principal components (which form the outlier map) indeed appear in clusters of high density.

While the DBSCAN algorithm requires some parameter choices (namely a search radius and the number of points required to form a dense region), it is often the case that the same parameter settings are useful over a range of scenarios provided that the data is consistently normalised.

## 3. Results and Discussion: Case Study

### 3.1.Short-term outliers

The time series data taken from the distributed control system in the VDU cover a range of dates from July 1, 2016 (00h:00m) until May 12, 2020 (21h:06m), 1,409 days in total. The frequency of the measurements is one reading per minute, which equates to 2,028,780 data points per sensor. For the 40 sensors available, the total dataset consists of 81,151,200 data points.

As described in *2.1. Statement of the problem* section, the most common method to determine outliers in the data is based on a $3\sigma$ criterion, using the mean and standard deviation of the entire signal. Standard deviation can be significantly affected by noise in the data and differ significantly across different sensors. For example, measurements that are more prone to variations caused by the process will often have a higher noise level. To quantify this effect, the signal-to-noise (STN) ratio was calculated and is presented in **Figure 6**.

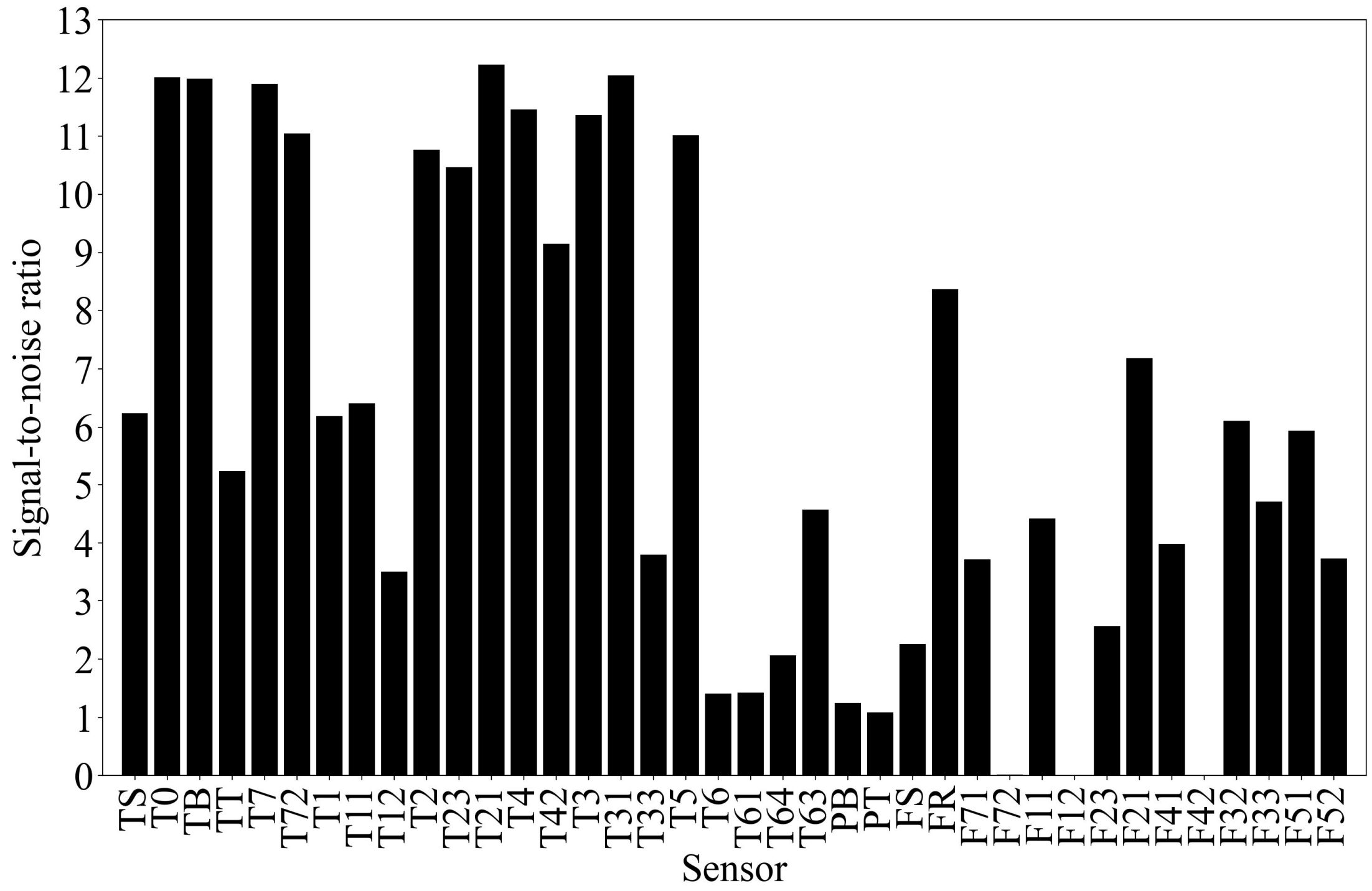

**Figure 6.** Signal-to-noise ratio of the 40 sensors computed as the ratio of mean and standard deviation of each signal.

Some of the measurements have very low signal-to-noise ratio values close to 0, namely F72, F12 and F42. The reason for these low values is that the noise level is artificially high due to some erroneous records of large magnitude which significantly deviate from the mean response. **Figure 7** shows such an example, signal F72, over the entire period.

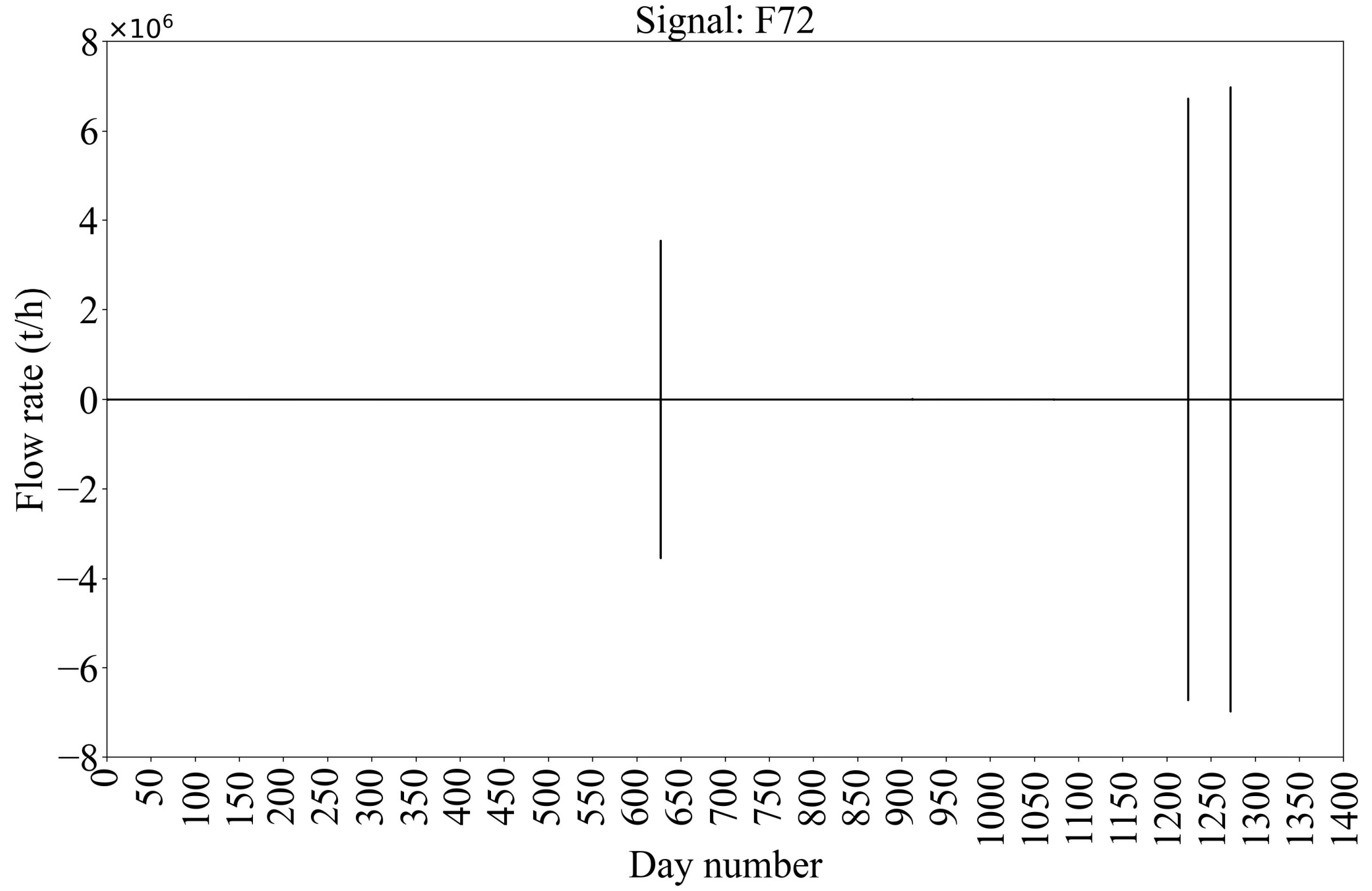

**Figure 7.** A visualisation of signal F72 with obvious outliers (*ca.* days 650, 1200, 1300) that cause a very low signal-to-noise ratio.

There are 6 significant outliers in the F72 signal on days around 650, 1200 and 1300 with values much higher than the overall mean response of the signal - the mean of the signal is 184 t/h, while the points identified above are up to *ca.* $3.8 \times 10^4$ times higher (the maximum value is 6,973,875 t/h). Hence, the large standard deviation of signal F72 (14481 t/h) produces a very low STN ratio.

Some sensors have relatively high signal-to-noise ratio (*e.g.* T0, TB, T7, *etc.*), while others have lower STN ratio (*e.g.* T6, T61, PB, *etc.*). Two examples of high and low signal-to-noise ratio are shown in **Figure 8** (more specifically signal T0 with STN=12.0 and signal T64 with STN=2.1).

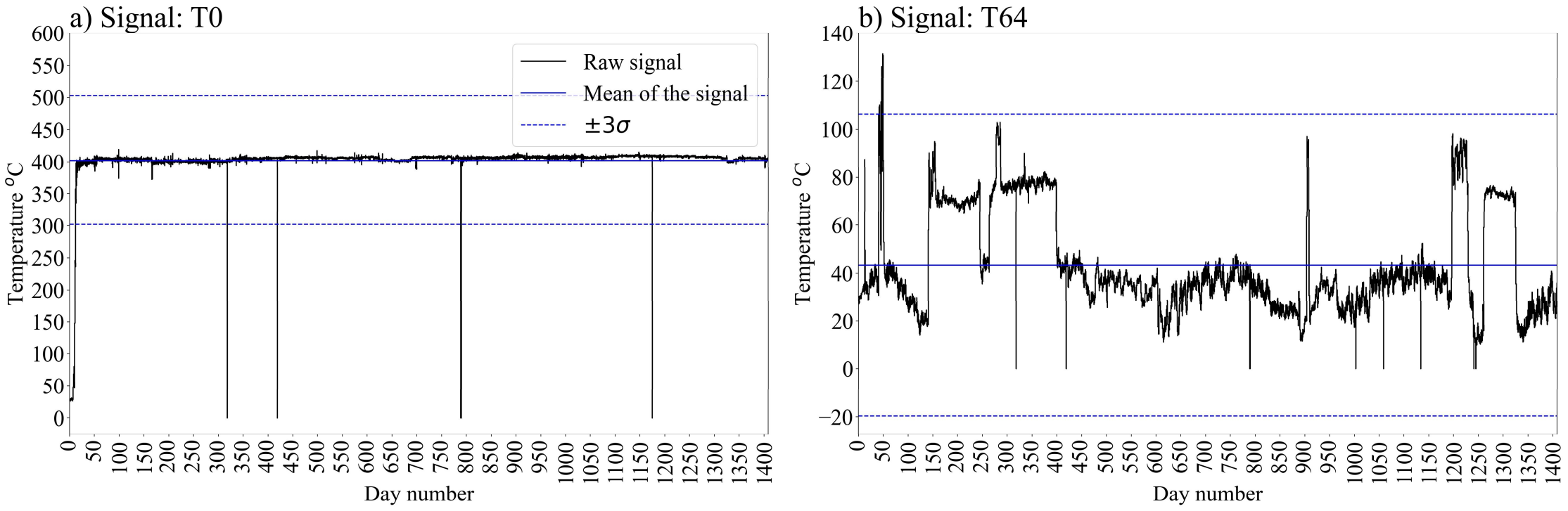

**Figure 8.** a) Signal T0 with high signal-to-noise ratio; b) Signal T64 with low signal-to-noise ratio. Calculated 3$\sigma$ thresholds applied to the entire time series are shown as dashed blue lines.

The 3$\sigma$ method calculated on the entire signal's data does not allow to identify the outliers accurately. To show this problem with 3$\sigma$ method, we discuss two signals: T0 and T64. The mean value of signal T0 is 402 °C, and 3$\sigma$ upper and lower thresholds are (301 – 503) °C. Every data point outside the upper and lower 3$\sigma$ bound is classed as an outlier. The reason for such a wide 3$\sigma$ range is the obvious outliers at days around (0, 300, 400, 800, 1150) which have values close to 0 °C. The points on these days are also the only outliers detected, as can be seen in **Figure 8a**. On the other hand, the mean value of signal T64 is 42 °C. It can be seen in **Figure 8b** that the same days clearly deviate from the overall trend in T64, although not as prominent as in signal T0. This is due to the relatively low mean value and higher variance of signal T64. The signal varies from approx. 15 °C to approx. 100 °C. This variance produces a high standard deviation (relative to the mean value) which then gives a wide 3$\sigma$ range. In the case of signal T64, 3$\sigma$ thresholds are -20 °C and 106 °C. Consequently, signal T64 is almost entirely within the 3$\sigma$ range and no outliers were identified, except for some spikes around day 50.

To further visualise the problem of misestimating the uncertainty range through calculation of 3$\sigma$ over the entire time series, signal T0 was shown between days 32 and 182 in **Figure 9**.

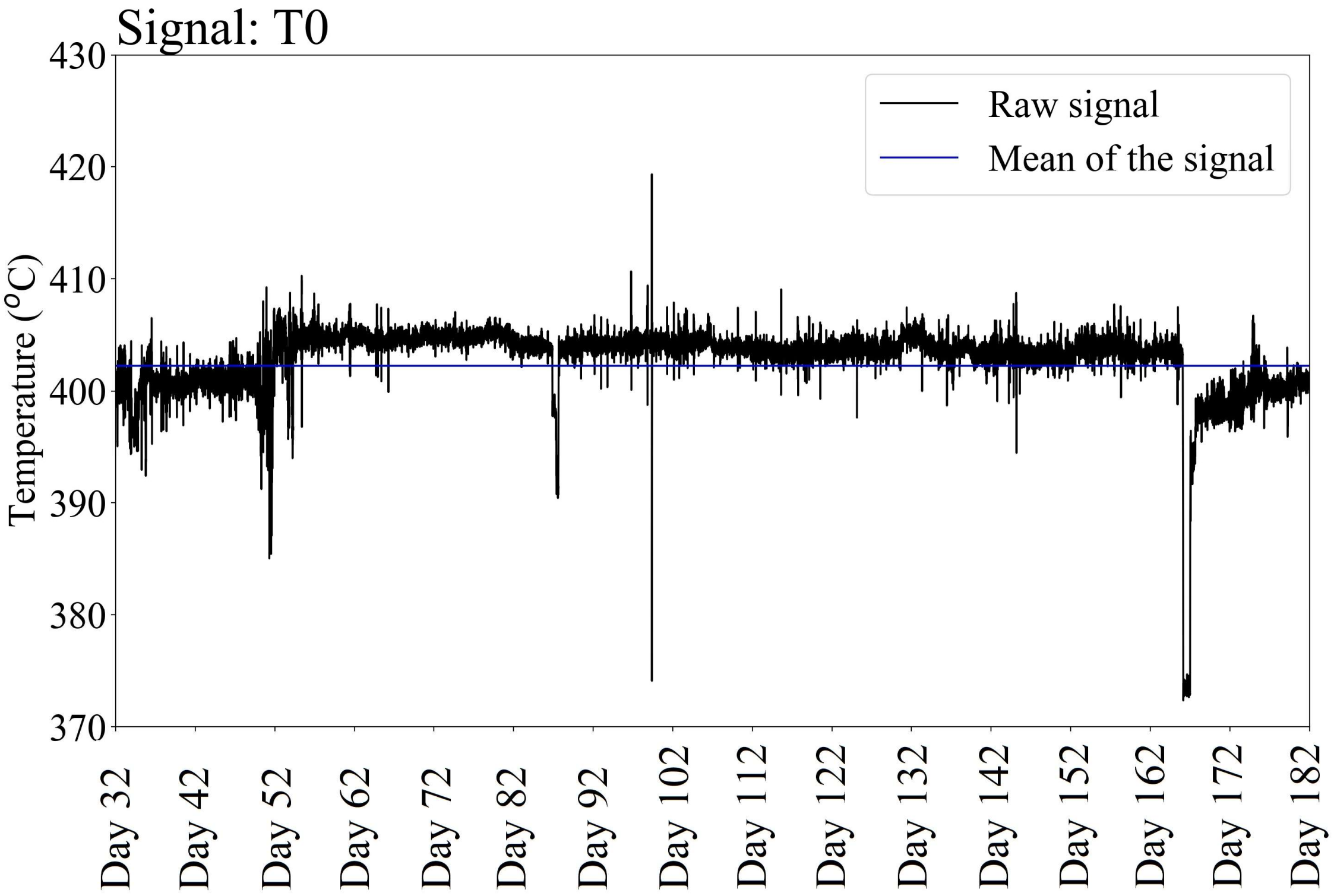

**Figure 9.** Signal T0 for days 32 to 182 showing the presence of outliers undetected when $3\sigma$ approach is applied to the entire time series (where $-3\sigma$ = 301 °C, and + $3\sigma$ = 503 °C).

It can clearly be seen in **Figure 9** that outliers are present, but as these lie within the upper and lower $3\sigma$ thresholds (301 – 503 °C), for example around days 102 or 142, they are not identified as such. $3\sigma$ is heavily affected by the variance in the data, amount and values of outliers (particularly the obvious outliers), making it difficult to detect less extreme outliers. Nevertheless, while one can expect such statistical measures to produce reliable results, a physicochemical understanding of the process and sensors under consideration is needed, *e.g.* it is unlikely that a temperature sensor with mean value of 402 °C measures real temperatures close to 0 °C; this unrealistic value suggests the sensor was malfunctioning or switched off. The traditional $3\sigma$ method applied to the entire time series works efficiently for obvious outliers, more specifically those that deviate significantly from the overall statistics, such as for signal T0.

Details for the modified method for piece-wise approximation of $3\sigma$ values (through change points detection) are discussed in *2.2. Short-term outliers* section, including one of the parameters required for the algorithm for change points detection, namely the minimum length of the piece $L_{min}$. Generally, this parameter defines the minimum length required between two change points. $L_{min}$ determines whether the length of the run is satisfactory, hence, it might affect the results of piece-wise approximated $3\sigma$ values. Thus, we study the effect of $L_{min}$ on outlier detection for signal T0 in **Figure 10**.

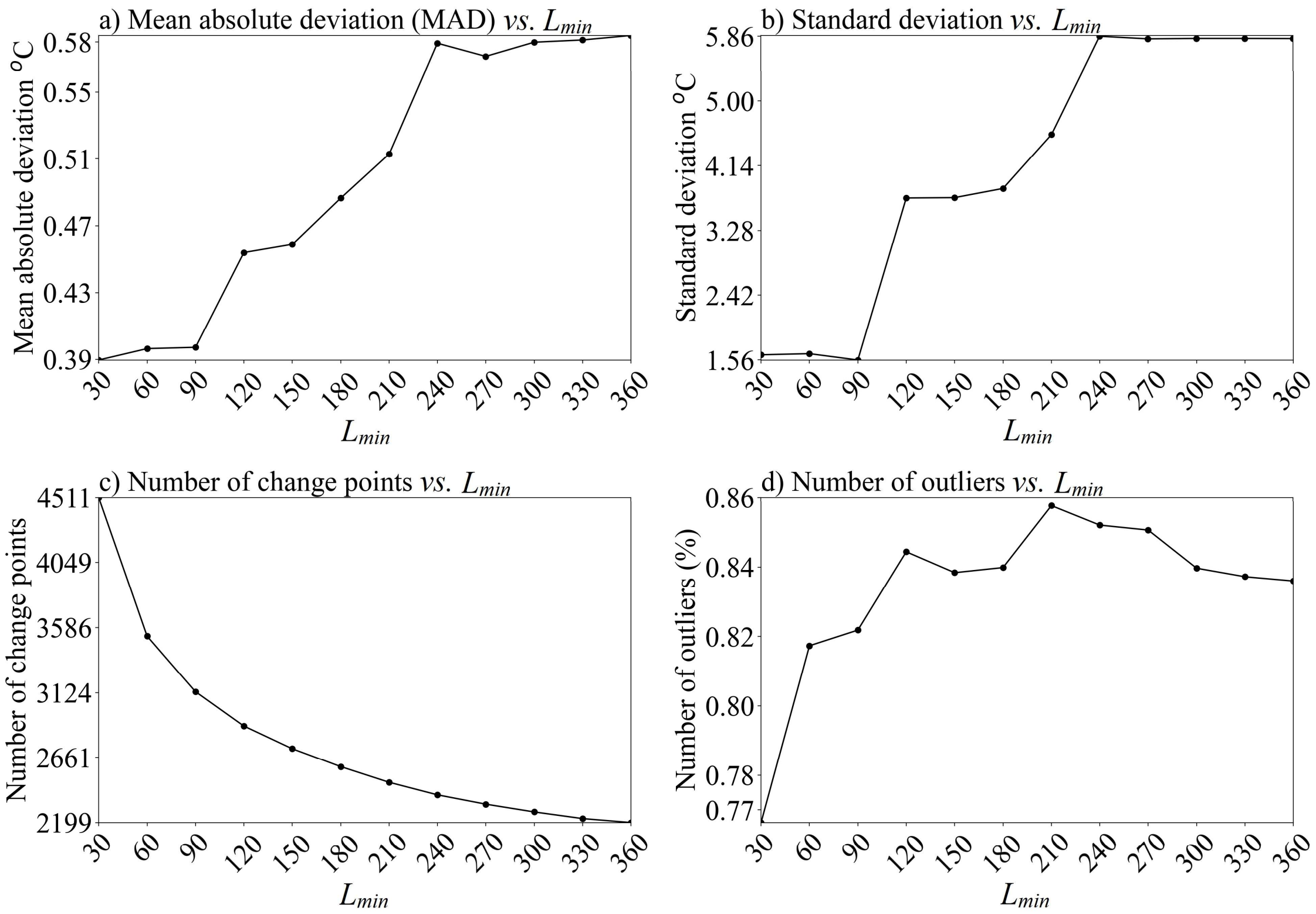

**Figure 10.** Effect of minimum length of the piece $L_{min}$ for change points detection on outliers detection for signal T0; a) Mean absolute deviation (MAD); b) Standard deviation; c) Number of change points; d) Number of outliers.

Mean absolute deviation (MAD) was calculated by subtracting the approximated piece-wise mean (calculated as a mean value between change points) from the raw signal, and then evaluating the absolute mean of that residue. **Figure 10a** shows the MAD for different $L_{min}$ values. 30 < $L_{min}$ < 90 produces the lowest MAD below 0.40 °C. There is a constant increase in the MAD in the range of 90 < $L_{min}$ < 240, while it plateaus from $L_{min}$ = 240 at *ca.* MAD = 0.58 °C. Similar relationships were observed for standard deviation (**Figure 10b**). Higher MAD and standard deviation indicate higher deviations between estimated piece-wise mean and the raw signal, which is an effect of longer pieces (greater distance between two change points). These longer pieces will include more data points which, consequently, contain more potential outliers, increasing the MAD and standard deviation. The effect of the longer pieces can also be seen from the number of change points detected (**Figure 10c**); more specifically, the number of change points detected decreases monotonically with increasing $L_{min}$. Fewer change points mean fewer approximated pieces, thus, longer pieces (and greater distances between change points). Finally, we calculated the number of outliers identified in relation to the total number of data points, 2,028,780 (**Figure 10d**). 0.77% was the lowest contribution of outliers detected for $L_{min} = 30$, the number of outliers identified for $60 < L_{min} < 90$ was around 0.82% of the total number of data, and for $L_{min} \geq 120$ the number of outliers does not change significantly (*ca.* 0.84% of all data for signal T0). The changes observed in the number of outliers indicate that there is a negligible effect of $L_{min}$ in the detection of outliers for $L_{min} > 60$. However, the number of outliers identified in relation to the total amount of data for signal T0 is relatively low (below 1%). In further analysis of the signals, $L_{min} = 60$ is used, as the MAD and standard deviation are the lowest for $30 < L_{min} < 60$, while there is a minimal effect beyond $L_{min} > 60$, for signal T0.

The results of outlier identification for signal T0 are shown in **Figure 11**, including piece-wise approximated mean and outlier-filtered data (the points identified as outliers replaced by the piece-wise approximated mean).

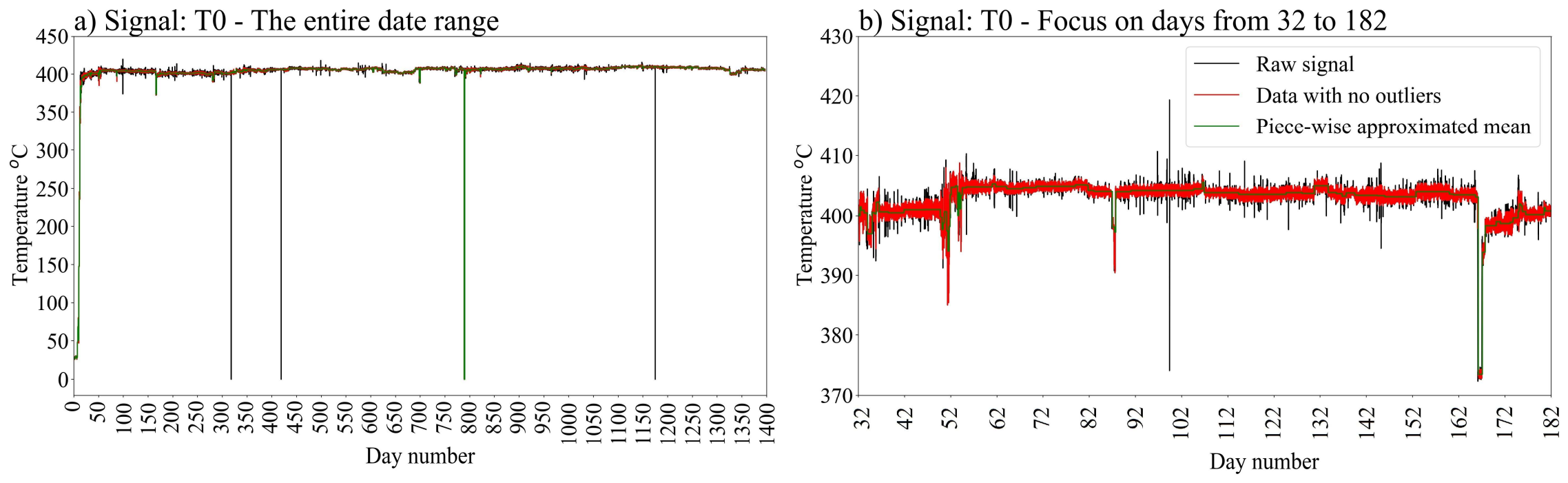

**Figure 11.** Outlier identification based on piece-wise mean approximation through change points detection for signal T0; **a)** general overview of the data, **b)** a focus on the period between days 32 and 182.

Firstly, the outlier detection based on piece-wise mean approximation through change points deals with the most obvious outliers in the data (**Figure 11a)**. Upon closer inspection of the signal T0 on days 32-182 (**Figure 11b)**, approximation of the mean between change points produces a baseline (green line) which is then used to calculate the upper and lower $3\sigma$ thresholds in a piece-wise manner. This approach allows changes in a mean response of the signal, while still detecting the short-term outliers reliably. In general, the number of short-term outliers was 16,582 for the entire signal T0 (which is equivalent to 0.82% of all data for signal T0). After filtering, the only obvious outliers still present are on days 1 – 13 and 789 – 791, as shown in **Figure 12**. These outliers have values very close to 0 °C which indicates malfunctioning of the sensor. The method based on piece-wise mean approximation was unable to detect these outliers because they last for longer periods (*i.e.* are long-term outliers).

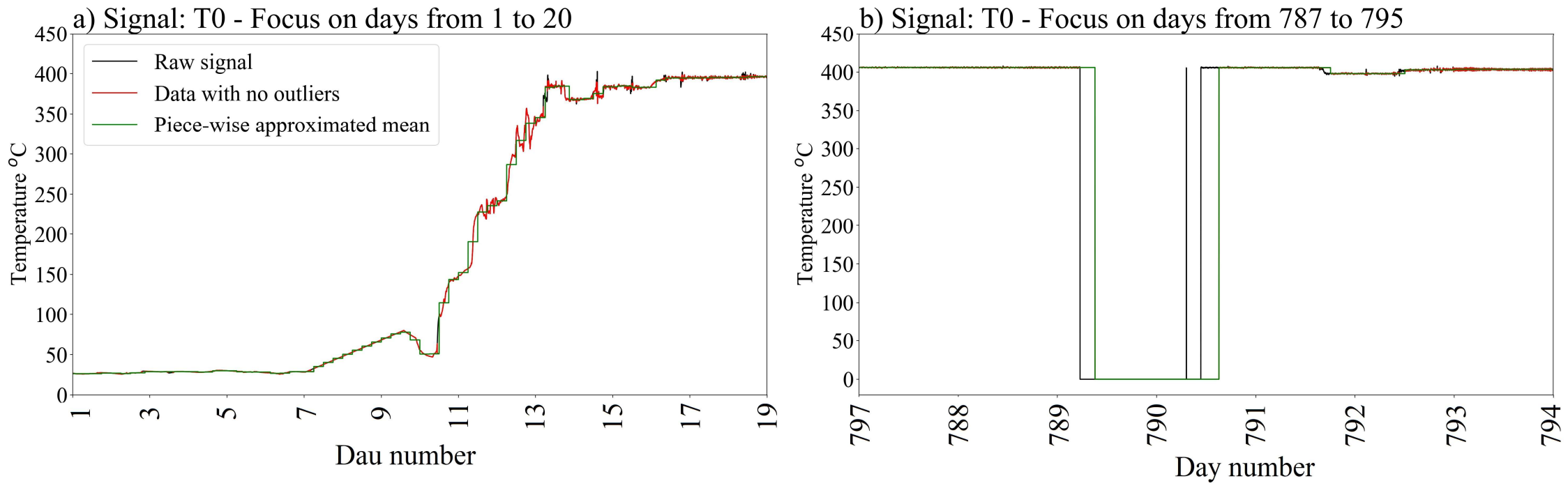

**Figure 12.** Long-term outliers on days 1 – 13 (a) and 789 - 791 (b), with piece-wise approximated mean (in green) and outliers-filtered data (in red).

The above considerations are based solely on signal T0. To showcase the high adaptability and capabilities of outlier detection with piece-wise mean approximation through change points, some of the other signals are shown, alongside the piece-wise approximated means and data with short-term outliers removed (**Figure 13**).

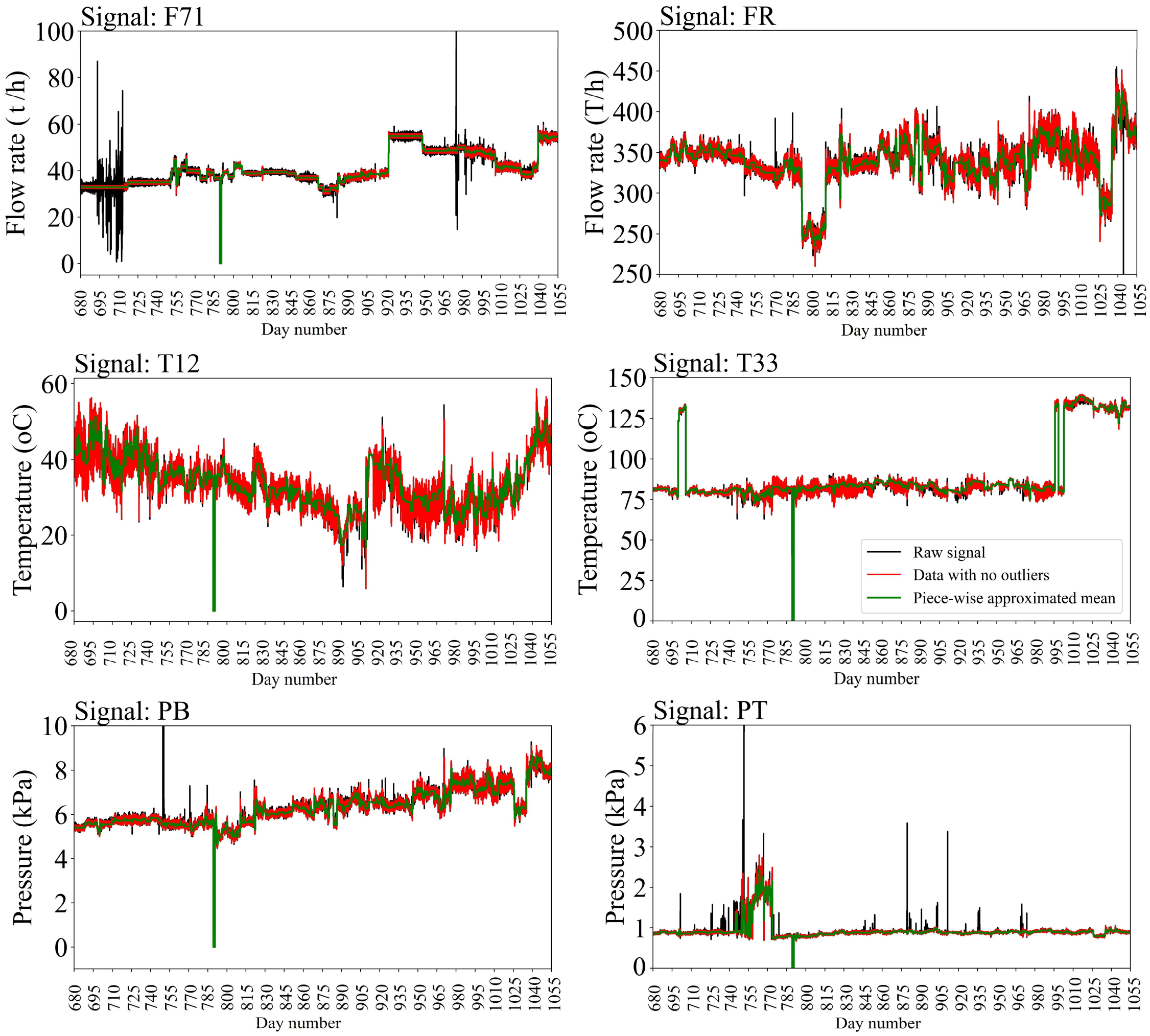

**Figure 13.** Other signals with their piece-wise approximated mean and outliers-filtered data.

Signal F71 has some abrupt changes to the readings between days 680 and 720 (values of *ca.* 0 - 90 t/h). The piece-wise 3$\sigma$ method successfully identified the records in this range as outliers. Signal FR and T12 have a highly variable noise level. Nevertheless, it was possible to calculate the piece-wise mean values to construct the corresponding 3$\sigma$ thresholds. Signal T33 has some increased values for longer periods at around 130 °C (instead of around 80 °C), for example days 695 – 710 and after day 1010. These changes in the operation of the sensor were reflected in the piece-wise mean which allows identification of outliers within these periods as well, despite having different statistical behaviour to the overall trend of the signal. The overall number of outliers is summarised in **Figure 14**. The lowest number of outliers detected (relative to all data) was 0.19% for F12 (equal to 3,854 data points) and the highest of 1.82% for F71 (equal to 36,720 points).

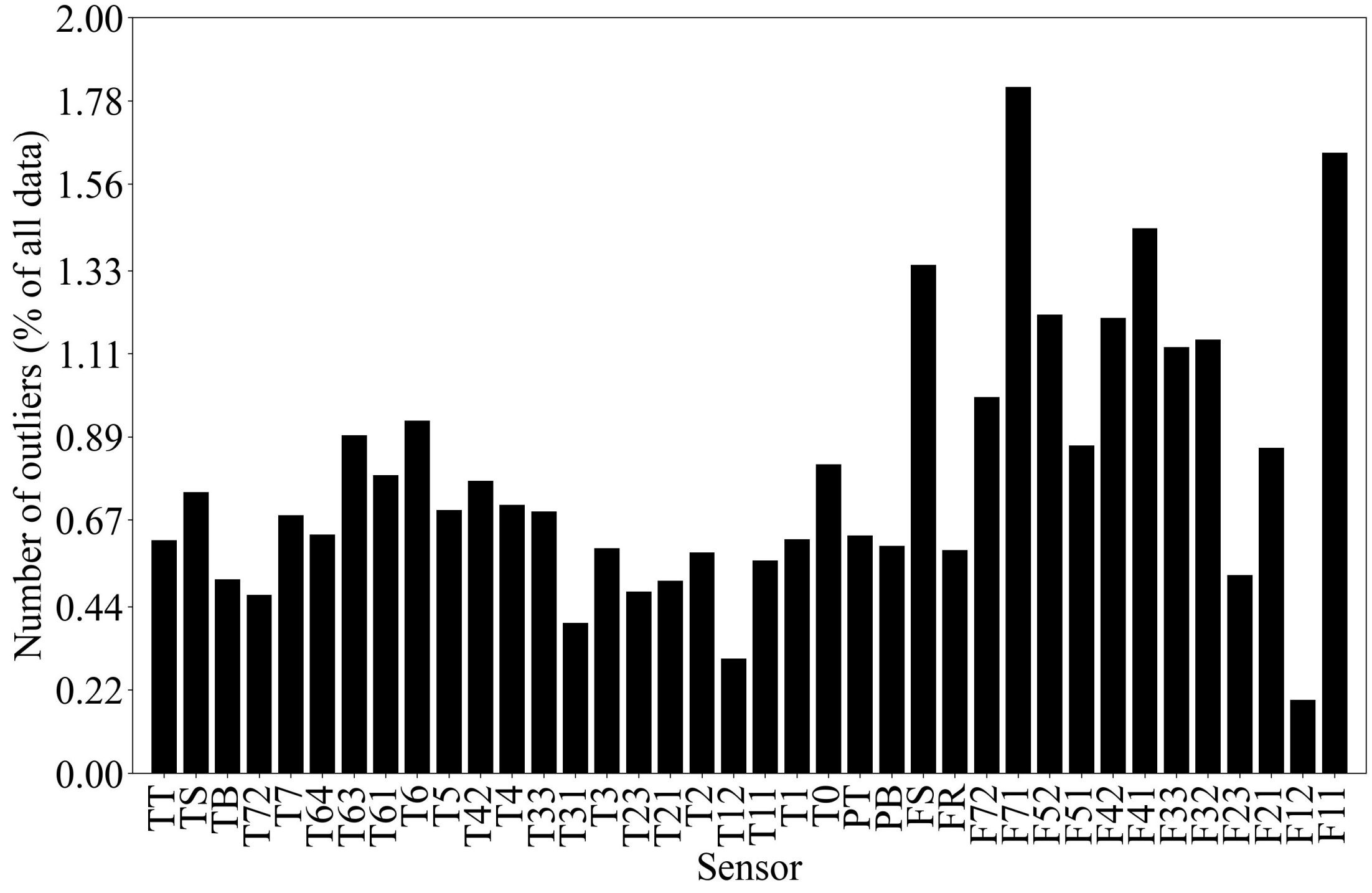

**Figure 14.** Number of outliers detected (fraction of data points identified as outliers).

### 3.2.Long-term outliers: Principal component analysis and DBSCAN clustering

The short-term outliers identified were replaced with the piece-wise approximated mean values before using PCA model. To ensure that each sensor has the same contribution to the PCA model, sensors data were scaled with their mean and unit variance before using the PCA model. The PCA model with up to 30 principal components was used to assess the impact of each principal component on the explained variance, reducing the number of features (*i.e.* sensors) from 40 sensors (23 for temperature, 2 for pressure and 15 for flow rate), as shown in **Figure 15**.

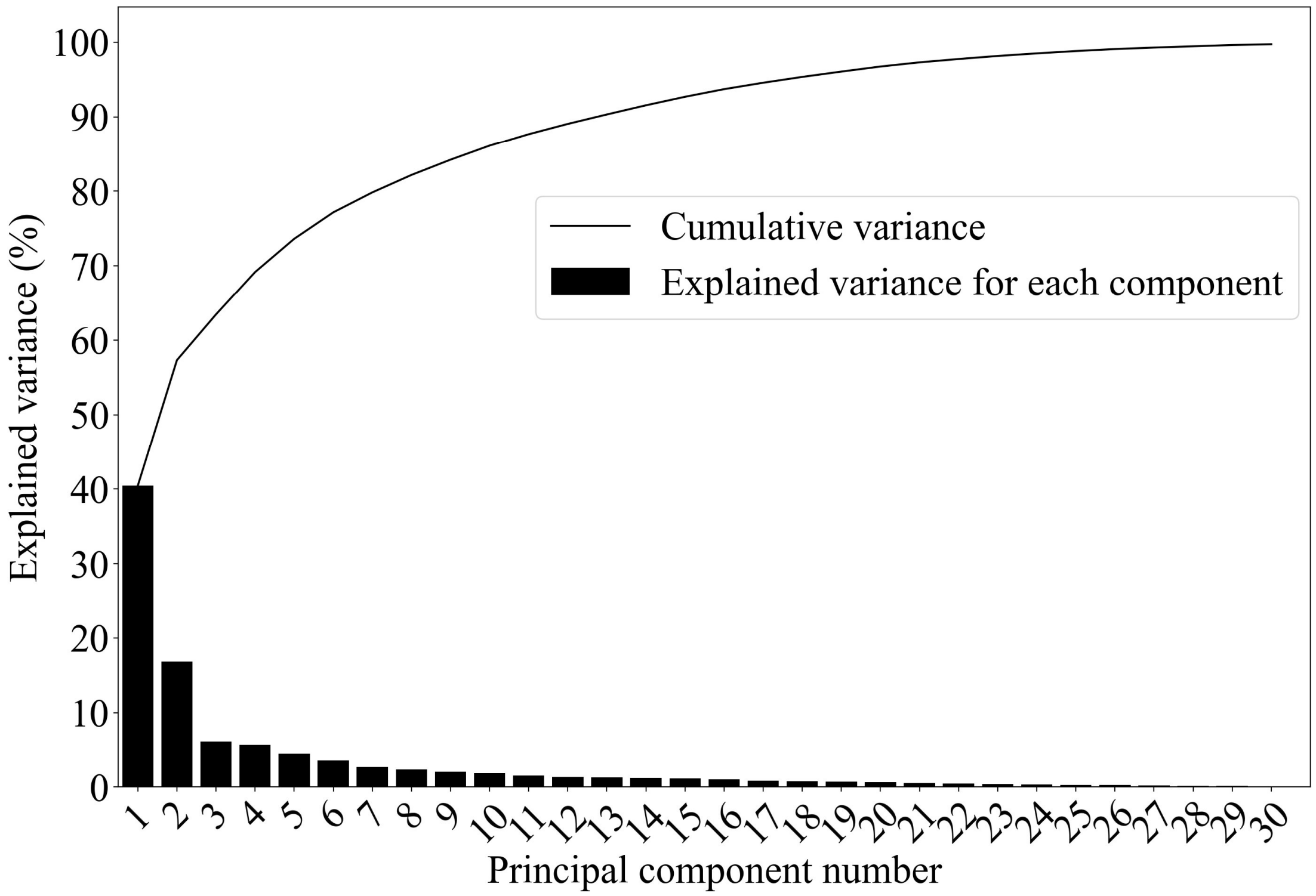

**Figure 15.** Explained variance through principal components (black line for cumulative variance and bars for each components variance).

The first most important principal component explains 40.5% of the variance, while the second principal component accounts for 16.8% of the explained variance. In total, the first two components explain *ca.* 57% of the variance. For the third and subsequent principal components, the explained variance decreases slowly. For example, 90% of the variance is explained by 13 principal components, while 95% and 99% of the variance is explained by 18 and 26 principal components, respectively. In terms of dimensionality reduction techniques, it is common practice to select the number of components to explain 90% of the variance; in this case, there is a three-fold reduction in dimensionality (from 40 to 13 components).[47,48]

To quantify the outliers, the first two principal components were used to calculate Hotelling's $T^2$ statistics (**Equation 11)**, shown in **Figure 16.**

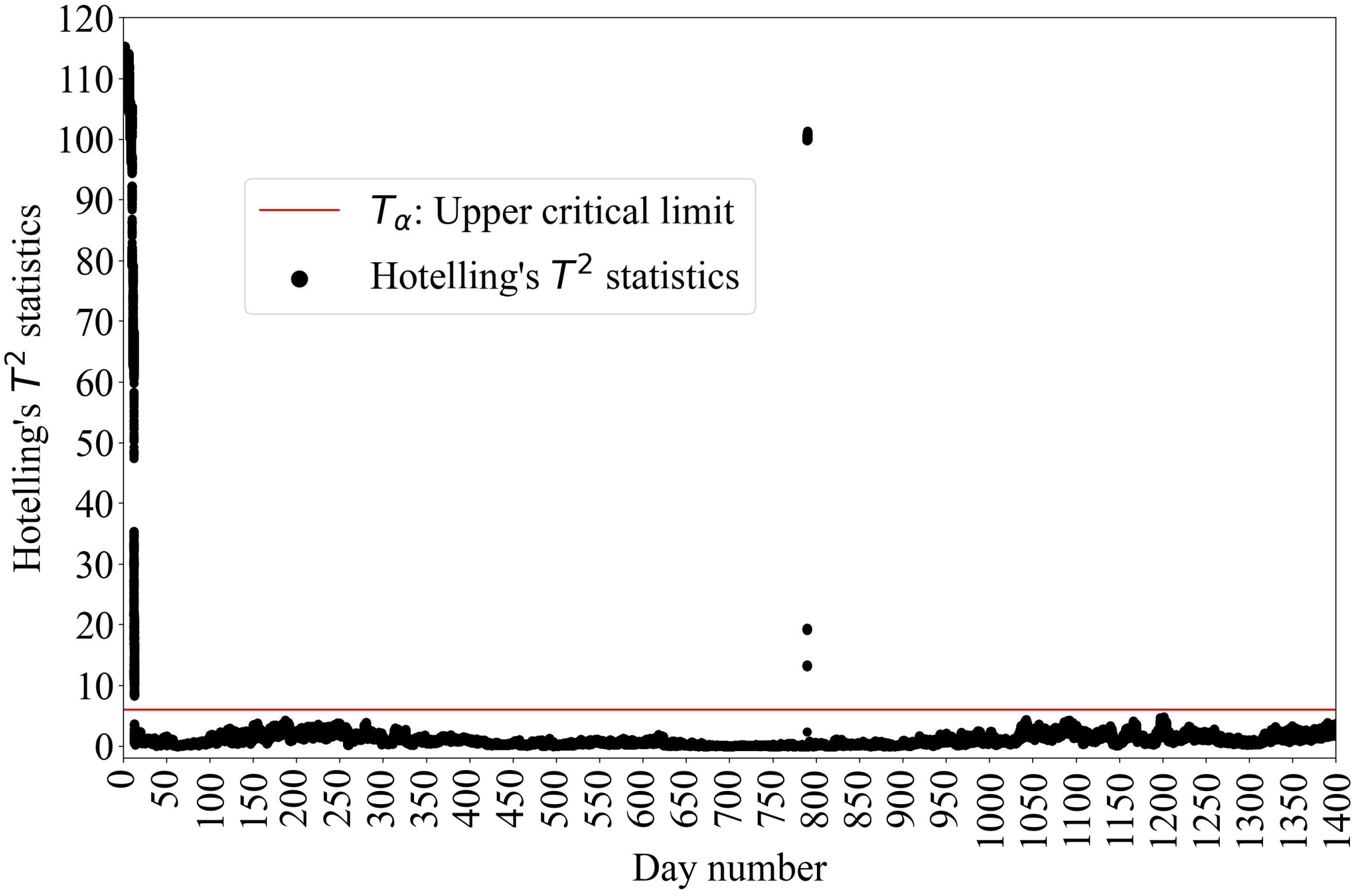

**Figure 16.** $T^2$ statistics used for long-term outliers identification.

The upper critical limit for the $T^2$ statistics was taken from F-distribution. Based on the degrees of freedom (2 principal components and 2,028,780 samples) and confidence level of $\alpha$=0.05, the upper critical limit is $T_\alpha$=5.99 (**Equation 12**). Values of $T^2$ above $T_\alpha$ ($T^2 > T_\alpha$) are classed as outliers, while those below $T_\alpha$ ($T^2 < T_\alpha$) are classed as non-outliers. There were 19,837 outliers identified in total. There are two main periods when $T^2 > T_\alpha$. The first one includes 17,820 data points, covering dates from day 1 (00:00) until day 13 (08:59). The second period includes 2,017 data points, covering dates from day 789 (06:06) until day 790 (15:05).

Higher values of $T^2$ statistics indicate a higher impact on the fraction of variance explained by principal components, in particular principal component 1 which explains the highest fraction of the variance (40.5%). Deviations (such as malfunction) in multiple sensors can increase the value of $T^2$. Some of the malfunctioning sensors are shown in **Figure 17**.

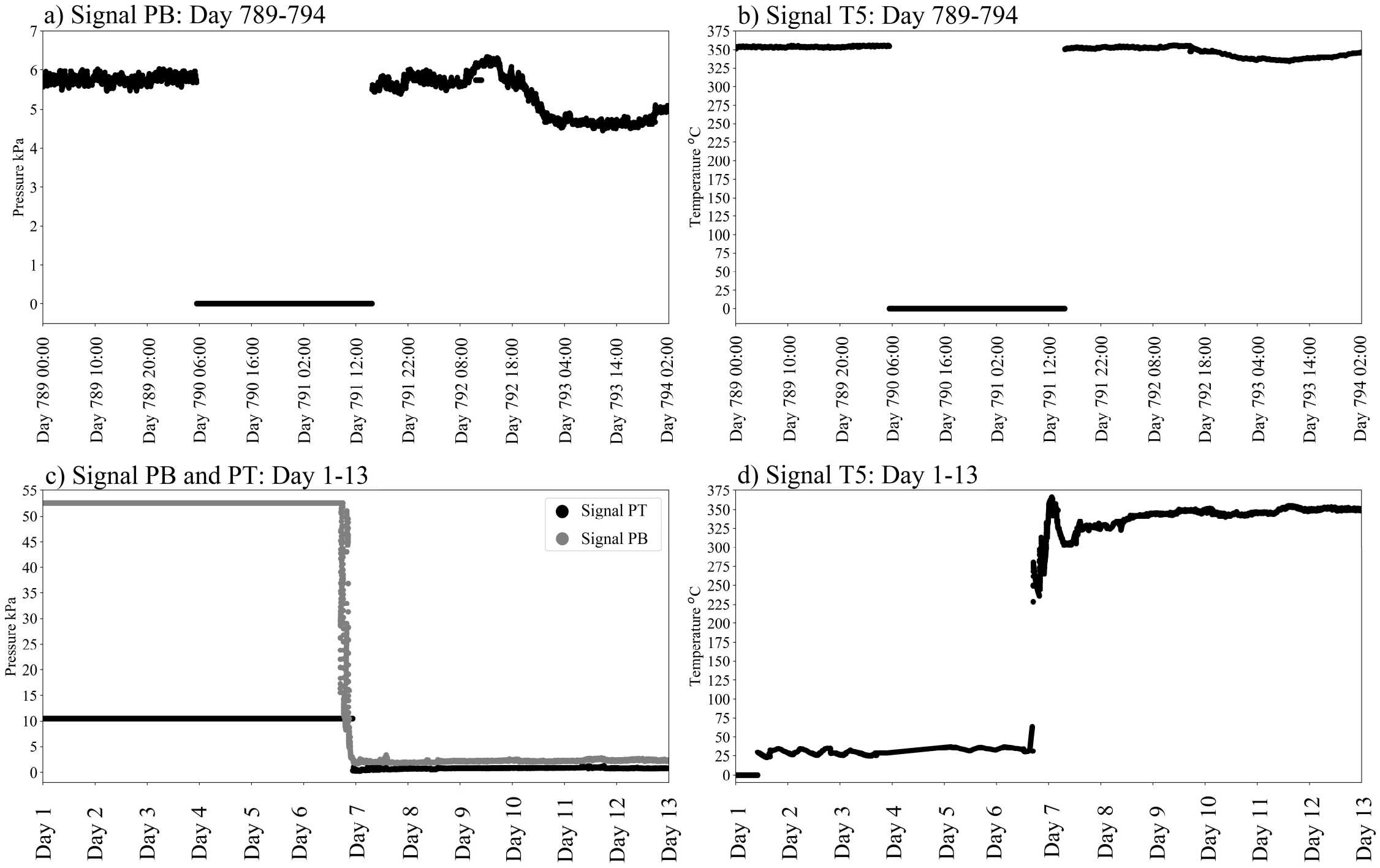

**Figure 17.** Examples of long-term outliers for multiple sensors: PB and T5 on days 789-794 (**a** and **b**); and TB, TP, T5 on days 1-13 (**c**-**d**).

During the second period (day 789-790), 34 sensors (out of 40) recorded values of 0 (°C, t/h or kPa, depending on the sensor), as shown in **Figure 17a-b**. During the first period (day 1 – 13), both pressure measurements were higher: *ca.* 52 and 11 kPa (for PB and PT), instead of 11 and 1 kPa from day 7. Some other sensors did not record 0 (°C or F/h) on day 1–7 but clearly malfunctioned (for example T5 recorded around 32 °C, while mean is 352 °C, as in **Figure 17d**). Hotelling's $T^2$ statistics has identified outliers which, on visual inspection, do indeed look like outliers. The possible reason for these long-term outliers is the abnormal functioning of the process, rather than single/multiple sensors. The first two principal components projection was used to construct an outlier map, shown in **Figure 18**.

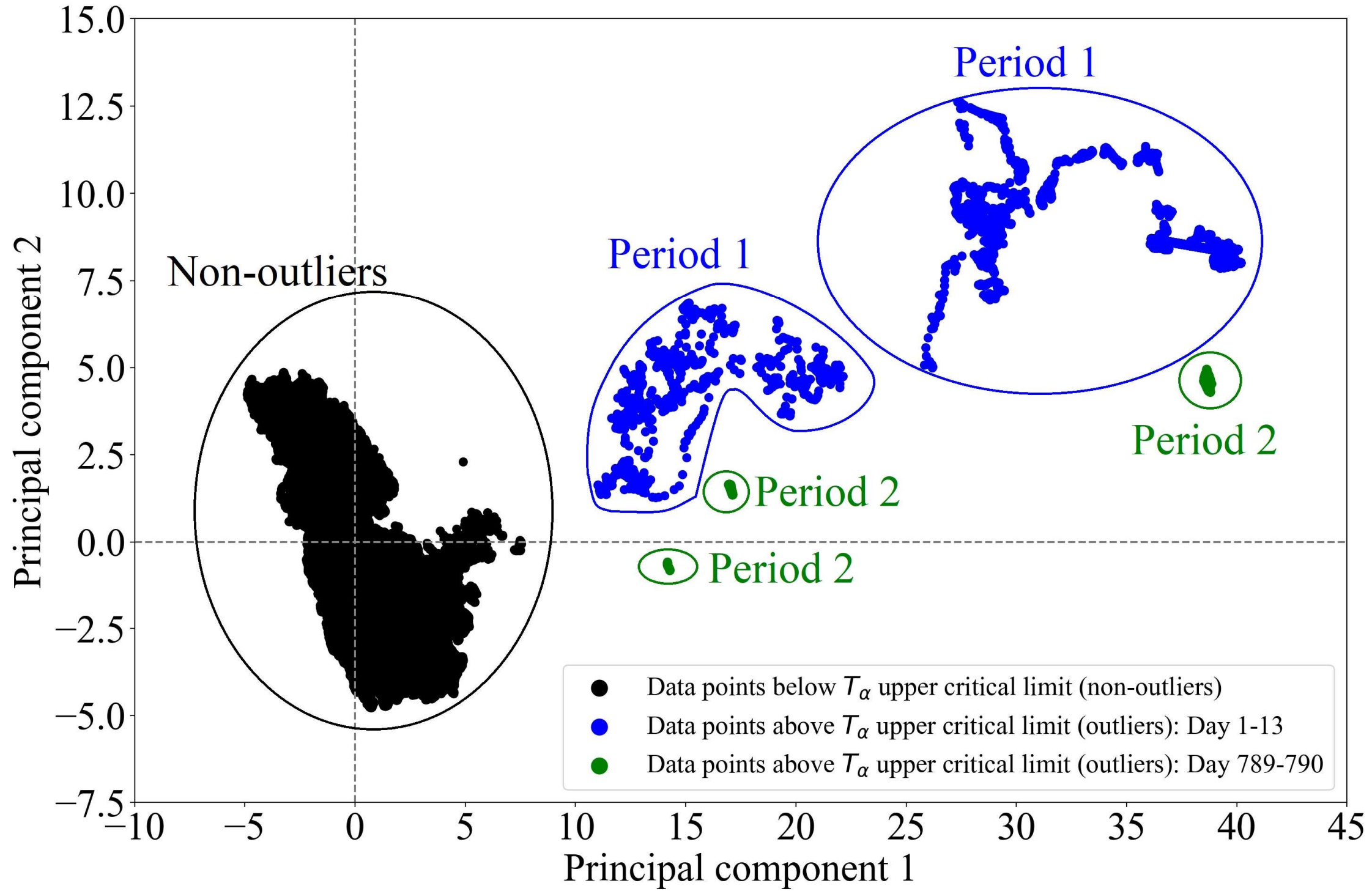

**Figure 18.** Outliers map showing long-term outliers in two periods: days 1-13 and 789-790.

The centre of non-outliers on the outlier map is positioned at (0, 0), due to the data centred with their mean. The range of principal components values covering the region of non-outliers is –4.9 < principal component 1 (x-axis) < 7.5 and –4.8 < principal component 2 (y-axis) < 12.6. All data points with principal component 1 (x-axis) > 7.5 are outliers. PCA has successfully identified the long-term outliers. The two periods of long-term outliers are clearly separated. Generally, period 1 (days 1-13) creates two clusters, while period 2 (days 789-790) creates three small clusters. To identify and classify these clusters, we used DBSCAN clustering method on the PCA results (outliers map).

We used the open-source algorithm available in scikit-learn package in Python. The minimum parameters required for the algorithm are epsilon $\varepsilon$ and so-called 'minimum samples in a neighbourhood'. Birant and Kut (**2007**) studied the effect of $\varepsilon$ and minimum samples in a neighbourhood on performance of DBSCAN and proposed a way to calculate $\varepsilon$ and minimum samples parameters.[49] Firstly, $\varepsilon$ can be calculated based on the *k*-distance method. We found that for the dataset in this work $\varepsilon = 0.7$ (while the default value in scikit-learn is $\varepsilon = 0.5$). Secondly, minimum samples in a neighbourhood can be taken as a logarithm of the number of samples which gives log(2,028,780)≈6 (while the default value in scikit-learn is 5). The outlier map with 8 clusters identified through DBSCAN is shown in **Figure 19**.

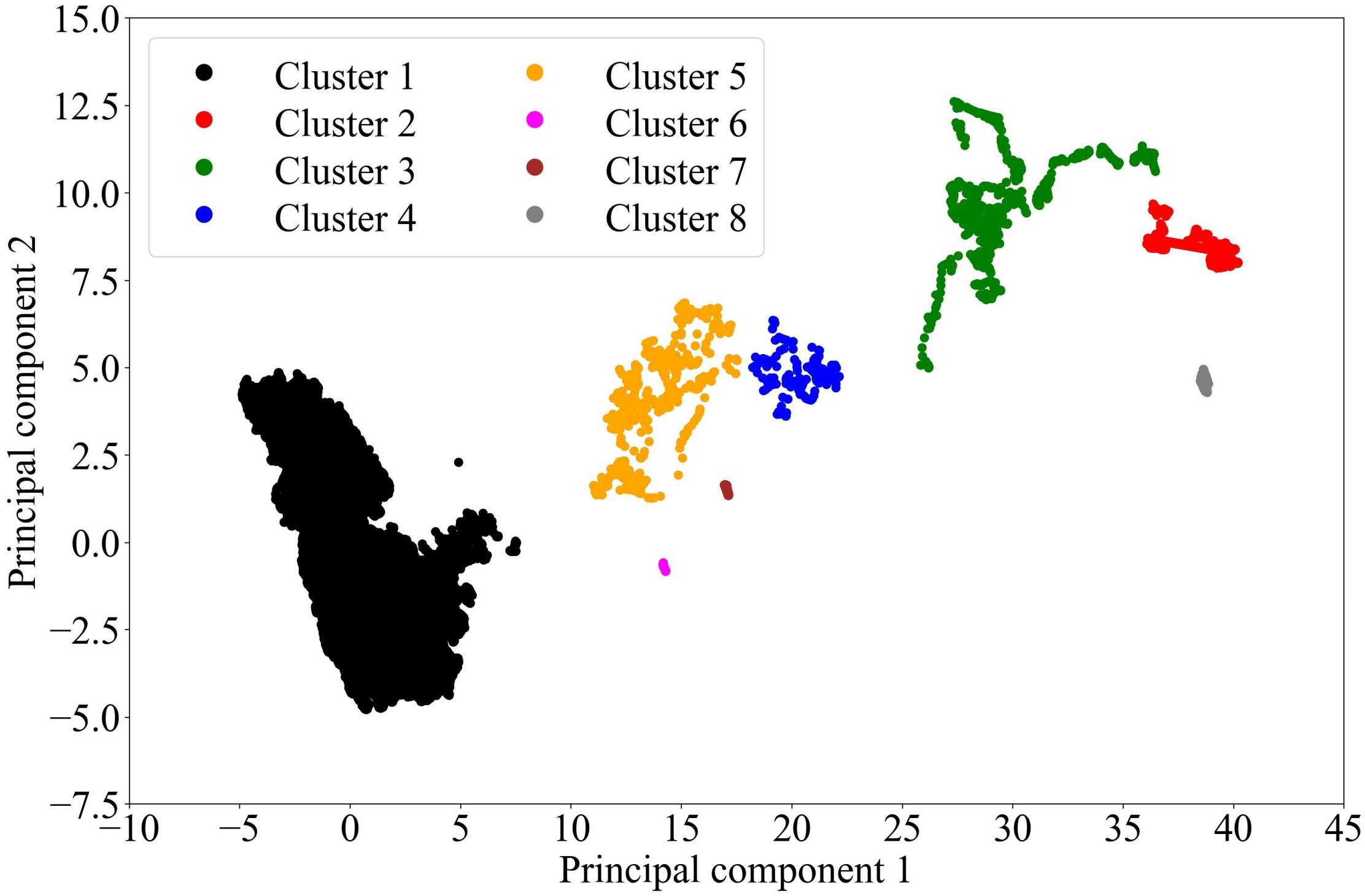

**Figure 19.** DBSCAN clustering results shown on the outlier map.

Cluster 1 contains all non-outlier data (2,008,943 in total). Clusters 2 to 5 identified days 1 – 13. Cluster 2 covers the widest range of outliers (starting from day 1 at 00:00 until day 10 at 02:59 (13,140 data points in total); Cluster 3 spans day 10 at 03:00 to day 12 at 17:59 (3,780 data points); Cluster 4 covers day 12 at 18:00 to day 12 at 20:59 (180 data points); Cluster 5 corresponds to day 12 at 21:00 to day 13 at 08:59 (719 data points). For the other period with long-term outliers (days 789-790), three further clusters were identified: cluster 6 on day 789 at 05:29 to day 789 at 06:05 (37 data points); Cluster 7 day 789 at 06:06 to day 789 at 09:05 (180 data points); and cluster 8 day 789 at 09:06 to day 790 at 15:05 (1,800 data points). Interestingly, the clusters were clearly separated into the two relevant periods. Also, the outliers were contiguous, *i.e.* going from one cluster to another alongside the duration of outlier period. The sensor data for these periods are coloured according to the corresponding cluster **Figure 20**.

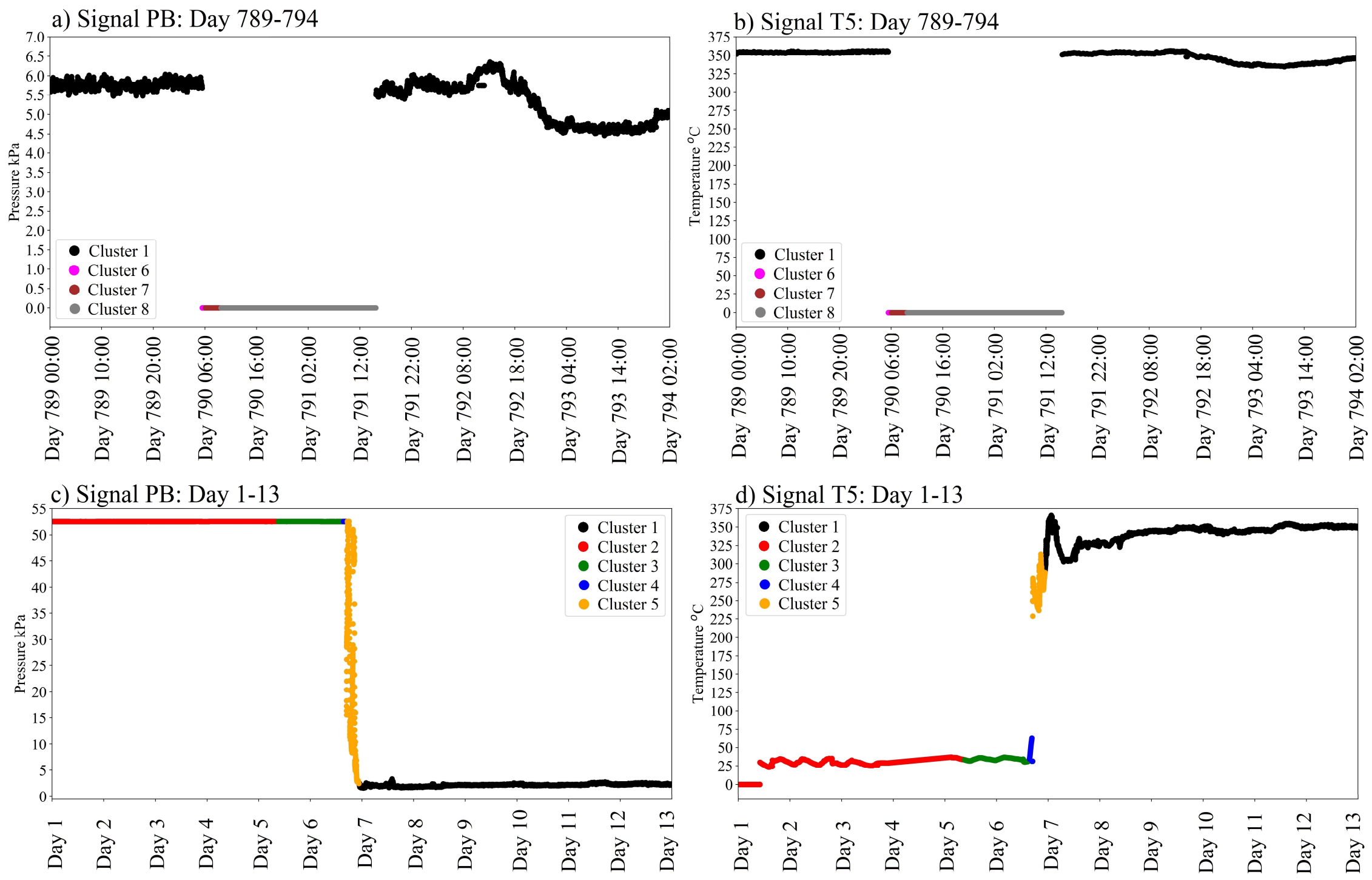

**Figure 20.** Long-term outliers in period 1 (days 1-13) and period 2 (days 789-790), with colours corresponding to the identified clusters.

**Figure 20** shows that period 2 (days 789-790) starts with cluster 6 followed by cluster 7 and then cluster 8, for both signals PB and T5 (**Figure 20a-b**). This pattern suggests that cluster 6 and 7 are a form of transition period within the outlier period, more specifically they are short-lived (cluster 6) and transition into other longer clusters (from 6 to 7 and from 7 to 8), which suggests that cluster 8 represents quasi-stationarity in the outlier period. Overall, this transition period would last as long as cluster 6 and 7 (3 hours and 37 minutes). Unfortunately, the data available for period 2 (days 1-13) are limited*, i.e.,* the long-term outlier (such as malfunctioning) is already present in the data which means that we are not able to recreate the entire outlier period. Nevertheless, the opposite pattern can be seen in period 1 to the one seen in period 2, more specifically, the so-called transition period is happening when the signal is going back to normal functioning (cluster 4 and 5), as shown in **Figure 20c-d**. Their combined duration of cluster 5 and 4 is 2 hours and 59 minutes, which is similar to the transition period in period 2. The outliers (which are visually obvious and are correctly detected by the PCA method) were correctly identified by DBSCAN which supports the consistency and accuracy of the PCA method.

## Conclusions

Data pre-treatment is a vital step to enhance the quality of data. Outlier detection is one of the steps in data pre-treatment. This work focused on detection of short- and long-term outliers in plant data (Vacuum Distillation Unit)

In the case of short-term outliers, we compared the most common method, namely $3\sigma$, with its modified version, *i.e.* piece-wise approximated $3\sigma$ determined by the change points identified on the basis of mean changes. It was shown that the traditional $3\sigma$ method applied to the entire time series produces unreliable results. More specifically, the values and number of outliers (particularly obvious outliers) can significantly shift the standard deviation of the signal which results in a misestimation of $3\sigma$ thresholds. The piece-wise approximated $3\sigma$ was shown to overcome this issue. However, it was impossible to detect long-term outliers with piece-wise approximated $3\sigma$ as these were identified as a quasi-steady state due to the reliance on change points detection driven by the mean response of the signal.

After the short-term outliers were identified and replaced with the piece-wise approximated mean values of the entire time series, we applied PCA to determine long-term outliers. Hotelling's $T^2$ statistics derived from the first two principal components was calculated and used to identify those long-term outliers. A combination of piece-wise approximated $3\sigma$ method with PCA provided a robust approach for both short- and long-term outliers detection. Furthermore, the first two principal components were used to construct so-called outlier maps and to identify clusters created through dimensionality reduction. It was found that long-term outliers create separated clusters from each other that can be identified reliably. The outliers (which were visually obvious and correctly detected by the PCA method) were also correctly identified by DBSCAN clustering which supported the consistency and accuracy of the PCA method. Unfortunately, only one of the two long-term outliers had start and end points in the available data, the other one was showing ongoing malfunctioning which limited the analysis of long-term outliers. We observed that the clusters in the outlier map represents the dynamics or transition states between correct functioning and malfunctioning of the sensors which provides important information for process control purposes.

## Acknowledgments

This work was supported by Innovate UK *via* a Knowledge Transfer Partnership (No. KTP10916) between the University of Manchester and Process Integration Limited.

## References

[1] D.M. Hawkins, Identification of Outliers. Springer, (1980).

[2] A. Blázquez-García, A. Conde, U. Mori, J.A. Lozano, A Review on outlier/Anomaly Detection in Time Series Data, ACM Comput. Surv. 54 (2021) 1–33.

[3] C.C. Aggarwal, Outlier analysis second edition, (2016).

[4] S. Basu, M. Meckesheimer, Automatic outlier detection for time series: an application to sensor data, Knowl. Inf. Syst. 11 (2007) 137–154.

[5] S. Mehrang, E. Helander, M. Pavel, A. Chieh, I. Korhonen, Outlier detection in weight time series of connected scales, in: 2015 IEEE Int. Conf. Bioinforma. Biomed., IEEE, 2015: pp. 1489–1496.

[6] J. Chen, W. Li, A. Lau, J. Cao, K. Wang, Automated load curve data cleansing in power systems, IEEE Trans. Smart Grid. 1 (2010) 213–221.

[7] K.M. Carter, W.W. Streilein, Probabilistic reasoning for streaming anomaly detection, in: 2012 IEEE Stat. Signal Process. Work., IEEE, 2012: pp. 377–380.

[8] A. Reddy, M. Ordway-West, M. Lee, M. Dugan, J. Whitney, R. Kahana, B. Ford, J. Muedsam, A. Henslee, M. Rao, Using gaussian mixture models to detect outliers in seasonal univariate network traffic, in: 2017 IEEE Secur. Priv. Work., IEEE, 2017: pp. 229–234.

[9] H.N. Akouemo, R.J. Povinelli, Time series outlier detection and imputation, in: 2014 IEEE PES Gen. Meet. Conf. Expo., IEEE, 2014: pp. 1–5.

[10] K. Hundman, V. Constantinou, C. Laporte, I. Colwell, T. Soderstrom, Detecting spacecraft anomalies using lstms and nonparametric dynamic thresholding, in: Proc. 24th ACM SIGKDD Int. Conf. Knowl. Discov. Data Min., 2018: pp. 387–395.

[11] S. Papadimitriou, J. Sun, C. Faloutsos, Streaming pattern discovery in multiple time-series, (2005).

[12] C. Shang, F. Yang, D. Huang, W. Lyu, Data-driven soft sensor development based on deep learning technique, J. Process Control. 24 (2014) 223–233.

[13] Ž. Ujević, I. Mohler, N. Bolf, Soft sensors for splitter product property estimation in CDU, Chem. Eng. Commun. 198 (2011) 1566–1578.

[14] Z.U. Andrijic, N. Bolf, Soft Sensors Application for Crude Distillation Unit Product Quality Estimation, Goriva i Maz. 50 (2011) 201.

[15] J.J. Macias-Hernandez, P. Angelov, X. Zhou, Soft sensor for predicting crude oil distillation side streams using evolving takagi-sugeno fuzzy models, in: 2007 IEEE Int. Conf. Syst. Man Cybern., IEEE, 2007: pp. 3305–3310.

[16] B. Bidar, M.M. Khalilipour, F. Shahraki, J. Sadeghi, A data-driven soft-sensor for monitoring ASTM-D86 of CDU side products using local instrumental variable (LIV) technique, J. Taiwan Inst. Chem. Eng. 84 (2018) 49–59.

[17] V. Gopakumar, S. Tiwari, I. Rahman, A deep learning based data driven soft sensor for bioprocesses, Biochem. Eng. J. 136 (2018) 28–39.

[18] A. Rogina, I. Šiško, I. Mohler, Ž. Ujević, N. Bolf, Soft sensor for continuous product quality estimation (in crude distillation unit), Chem. Eng. Res. Des. 89 (2011) 2070–2077.

[19] J.W. Osborne, A. Overbay, The power of outliers (and why researchers should always check for them), Pract. Assessment, Res. Eval. 9 (2004) 6.

[20] M. Daszykowski, K. Kaczmarek, Y. Vander Heyden, B. Walczak, Robust statistics in data analysis—A review: Basic concepts, Chemom. Intell. Lab. Syst. 85 (2007) 203–219.

[21] M. Novak, I. Mohler, M. Golob, Ž.U. Andrijić, N. Bolf, Continuous estimation of kerosene cold filter plugging point using soft sensors, Fuel Process. Technol. 113 (2013) 8–19.

[22] D. Wang, J. Liu, R. Srinivasan, Data-driven soft sensor approach for quality prediction in a refining process, IEEE Trans. Ind. Informatics. 6 (2009) 11–17.

[23] S. Aminikhanghahi, D.J. Cook, A survey of methods for time series change point detection, Knowl. Inf. Syst. 51 (2017) 339–367.

[24] T. Pohlert, Non-parametric trend tests and change-point detection, CC BY-ND. 4 (2016).

[25] A.S. Polunchenko, A.G. Tartakovsky, State-of-the-art in sequential change-point detection, Methodol. Comput. Appl. Probab. 14 (2012) 649–684.

[26] J. Chen, A.K. Gupta, On change point detection and estimation, Commun. Stat. Comput.

30 (2001) 665–697.

[27] Y. Zheng, L. Liu, L. Wang, X. Xie, Learning transportation mode from raw gps data for geographic applications on the web, in: Proc. 17th Int. Conf. World Wide Web, 2008: pp. 247–256.

[28] L. Wei, E. Keogh, Semi-supervised time series classification, in: Proc. 12th ACM SIGKDD Int. Conf. Knowl. Discov. Data Min., 2006: pp. 748–753.

[29] S. Reddy, M. Mun, J. Burke, D. Estrin, M. Hansen, M. Srivastava, Using mobile phones to determine transportation modes, ACM Trans. Sens. Networks. 6 (2010) 1–27.

[30] E. Keogh, S. Chu, D. Hart, M. Pazzani, An online algorithm for segmenting time series, in: Proc. 2001 IEEE Int. Conf. Data Min., IEEE, 2001: pp. 289–296.

[31] L. Lacasa, B. Luque, F. Ballesteros, J. Luque, J.C. Nuno, From time series to complex networks: The visibility graph, Proc. Natl. Acad. Sci. 105 (2008) 4972–4975.

[32] V. Moskvina, A. Zhigljavsky, An algorithm based on singular spectrum analysis for change-point detection, Commun. Stat. Comput. 32 (2003) 319–352.

[33] C. Alippi, G. Boracchi, D. Carrera, M. Roveri, Change detection in multivariate datastreams: Likelihood and detectability loss, ArXiv Prepr. ArXiv1510.04850. (2015).

[34] D.J. Cook, N.C. Krishnan, Activity learning: discovering, recognizing, and predicting human behavior from sensor data, John Wiley & Sons, 2015.

[35] I. Cleland, M. Han, C. Nugent, H. Lee, S. McClean, S. Zhang, S. Lee, Evaluation of prompted annotation of activity data recorded from a smart phone, Sensors. 14 (2014) 15861–15879.

[36] V. Chandola, R.R. Vatsavai, Scalable Time Series Change Detection for Biomass Monitoring Using Gaussian Process., in: CIDU, 2010: pp. 69–82.

[37] P.K. Bhattacharya, Maximum likelihood estimation of a change-point in the distribution of independent random variables: general multiparameter case, J. Multivar. Anal. 23 (1987) 183–208.

[38] C. Zou, G. Yin, L. Feng, Z. Wang, Nonparametric maximum likelihood approach to multiple change-point problems, Ann. Stat. 42 (2014) 970–1002.

[39] I.J. Myung, Tutorial on maximum likelihood estimation, J. Math. Psychol. 47 (2003)

90–100.

[40] R. Killick, I. Eckley, changepoint: An R package for changepoint analysis, J. Stat. Softw. 58 (2014) 1–19.

[41] J. Plasse, H. Hoeltgebaum, N.M. Adams, Streaming changepoint detection for transition matrices, Data Min. Knowl. Discov. (2021) 1–30.

[42] F. Pedregosa, G. Varoquaux, A. Gramfort, V. Michel, B. Thirion, O. Grisel, M. Blondel, P. Prettenhofer, R. Weiss, V. Dubourg, Scikit-learn: Machine learning in Python, J. Mach. Learn. Res. 12 (2011) 2825–2830.

[43] M. Hubert, P. Rousseeuw, T. Verdonck, Robust PCA for skewed data and its outlier map, Comput. Stat. Data Anal. 53 (2009) 2264–2274.

[44] M. Hubert, S. Engelen, Robust PCA and classification in biosciences, Bioinformatics. 20 (2004) 1728–1736.

[45] J.P. George, Z. Chen, P. Shaw, Fault detection of drinking water treatment process using PCA and Hotelling's T2 chart, World Acad. Sci. Eng. Technol. 50 (2009) 970–975.

[46] M. Ester, H.-P. Kriegel, J. Sander, X. Xu, A density-based algorithm for discovering clusters in large spatial databases with noise., in: Kdd, 1996: pp. 226–231.

[47] B. Lin, B. Recke, J.K.H. Knudsen, S.B. Jørgensen, A systematic approach for soft sensor development, Comput. Chem. Eng. 31 (2007) 419–425.

[48] X. Yu, P. Chum, K.-B. Sim, Analysis the effect of PCA for feature reduction in non-stationary EEG based motor imagery of BCI system, Optik (Stuttg). 125 (2014) 1498–1502.

[49] D. Birant, A. Kut, ST-DBSCAN: An algorithm for clustering spatial–temporal data, Data Knowl. Eng. 60 (2007) 208–221.